\documentclass[11pt,a4paper]{article}
\usepackage{axodraw}
\usepackage{epsfig}
\usepackage{epic}
\usepackage{float}
\usepackage{afterpage}
\usepackage{amsmath}
\renewcommand{\baselinestretch}{1.2}\normalsize

\newcommand{\be}{\begin{equation}}
\newcommand{\ee}{\end{equation}}

\newcommand{\beq}{\begin{equation}}
\newcommand{\eeq}{\end{equation}}
\def\intg4#1{\int \! \frac{d^4 #1}{(2 \pi)^4}}
\def\slash#1{#1\!\!\!/\!\,\,}

\newcommand{\bea}{\begin{eqnarray}}
\newcommand{\eea}{\end{eqnarray}}
\newcommand{\klgl}{\:\hbox to -0.2pt{\lower2.5pt\hbox{$\sim$}\hss}
{\raise3pt\hbox{$<$}}\:}
\newcommand{\grgl}{\:\hbox to -0.2pt{\lower2.5pt\hbox{$\sim$}\hss}
{\raise3pt\hbox{$>$}}\:}
\begin{document}
%
%
\markboth{ }{ }
\renewcommand{\baselinestretch}{1.1}\normalsize
\vspace*{-0.5cm}
\hfill TUM-HEP-364/99

\vspace*{2cm}
\bigskip
\bigskip
\begin{center}
{\huge\bf{The Thermal Renormalization Group for Fermions, Universality, and the
Chiral Phase-Transition}} 
\end{center}
\bigskip
{\begin{center}
\vspace*{0.5cm}
   {\begin{center} {\large\sc
                Bastian Bergerhoff\footnote{\makebox[1.cm]{Email:}
                                  Bastian.Bergerhoff@Physik.TU-Muenchen.DE},
                Johannes Manus\footnote{\makebox[1.cm]{Email:}
                                  Johannes.Manus@Physik.TU-Muenchen.DE} \\
                and J\"urgen Reingruber\footnote{\makebox[1.cm]{Email:}
                                  Juergen.Reingruber@Physik.TU-Muenchen.DE}}
    \end{center} }
\vspace*{0cm} {\it \begin{center}
    Institut f\"ur Theoretische Physik, Technische
    Universit\"at M\"unchen,          \\
    James--Franck--Stra\ss e, D--85748 Garching, Germany
    \end{center} }
\vspace*{1cm}
\end{center}}
\setcounter{footnote}{0}
\bigskip
\vspace*{1cm}\begin{abstract}
\noindent
We formulate the thermal renormalization group, an implementation of the
Wilsonian RG in the real-time (CTP) formulation of finite temperature field
theory, for fermionic fields.
Using a model with scalar and fermionic degrees of freedom which should
describe the two-flavor chiral phase-transition, we discuss the mechanism
behind fermion decoupling and universality at second order transitions.
It turns out that an effective mass-like term in the fermion propagator which
is due to thermal fluctuations and does not break chiral symmetry is necessary
for fermion decoupling to work. 
This situation is in contrast to the high-temperature limit, where the
dominance of scalar over fermionic degrees of freedom is due to the different
behavior of the distribution functions.
The mass-like contribution is the leading thermal effect in the fermionic 
sector and is missed if a derivative expansion of the fermionic propagator 
is performed.
We also discuss results on the phase-transition of the model considered
where we find good agreement with results from other methods.
\end{abstract}

\newpage

\renewcommand{\baselinestretch}{1.2}\normalsize
\section{Introduction}

Finite temperature field theory has become a mature part of theoretical
particle physics over the past 25 years. 
Since the seminal work of Kirzhnits and Linde 
\cite{KirLind}
on symmetry restoration at high temperatures there has been a strong
theoretical interest in methods and results in the field.
In recent years, the experimental quest for the high temperature state of
strongly interacting matter, the quark-gluon plasma, has attracted additional
interest.
Nevertheless, there is a large number of open problems both on a
phenomenological as well as on a fundamental level.
These range from the equation of state of nuclear matter
at high temperature and density (of interest e.g. in astrophysics) to questions
about the approach of a quantum system to thermal equilibrium.

The most general approach to quantum field theory in a nontrivial "environment"
(as specified by a density matrix at some initial time $t_i$) was proposed
by Schwinger
\cite{Schwinger}
and Keldysh
\cite{Keldysh} 
and is known under the name of the closed time-path (CTP-)formulation of
quantum field theory.
This approach allows for the calculation of operator expectation values in quite
general systems, including systems in thermal and chemical equilibrium.
The special case of equilibrium physics has, because of its relative
simplicity, attracted most of the attention in the field.
It has also led to two (in principle equivalent) formulations, known as the
Matsubara- or imaginary-time and the real-time formulation of thermal field
theory (see e.g.~\cite{TFT,LeBellac}).
As the names suggest, in the (more widely used) Matsubara-approach one
formulates the theory in Euclidean space (with a finite time-interval), the
price to pay for this being the necessity of analytically continuing all
time-arguments of Green-functions at the end of the calculations.
The real-time formulation, on the other hand, works in Minkowski space
throughout and thus the need to do analytical continuations never arises.

As long as one is mainly interested in static quantities like the free energy
etc, there is of course no need to consider time-dependence. 
Since both formulations should yield the same answers for all quantities one
will usually pick the technically simpler one -- which most of the time will be
the Matsubara approach. 
However, even in a calculation within a thermal system, time-dependent
correlation functions carry important physical information to be interpreted
within linear response theory
\cite{LeBellac}.
Thus for example the decay of out-of-equilibrium fluctuations of a given system
is governed by the "damping rate", being connected to the imaginary part of the
self energy of the corresponding fields.
This quantity plays an important r{\^o}le in systems approaching the critical
temperature of a second  order phase-transition. 
In particular it encodes the phenomenon of critical slowing down
\cite{CSD}
-- the fact that the equilibration time diverges as $T \rightarrow T_c$ -- and
governs the initial density of possible topological defects
\cite{Zurek}.
It also provides a link between fundamental physics and a commonly used
phenomenological approach to out-of-equilibrium systems using
Langevin-equations
\cite{Langevin1,Langevin2,Langevin3}
with potentially large qualitative impact on the behavior of such systems
\cite{BMM}.
The fact that realistic systems will be prepared at some temperature and then
cooled (or heated) "through" a phase-transition during some finite time thus
provides a strong motivation to study thermal field theory using the real-time
formulation.

As for quantum field theory at vanishing temperature, there are in general no
exact analytical methods to do calculations in finite temperature field theory.
On the other hand, the methods we use in standard field theory may be more or
less trivially adapted to systems in or close to thermal equilibrium. 
One may in particular tackle a number of nontrivial problems using standard
perturbation theory, i.e.~the expansion in some small coupling.
It is however widely known that due to the emergence of an additional
dimensionful parameter -- namely the temperature $T$ -- the usual criteria for
the (superficial) convergence of perturbation theory are modified.
It turns out that even theories which are weakly coupled at $T\!=\!0$ may
not be correctly described in perturbation theory close to a second or
weakly first order phase-transition. 
The reason is that at a critical point the correlation length of the order
parameter field diverges and the theory is correlated over all scales.
In such a situation the behavior of the perturbative series is drastically
modified by infrared phenomena.
Thus for example perturbation theory predicts a first order phase-transition
for $O(N)$-symmetric scalar theories in $3+1$ dimension, whereas we know this
transition to be of second order
\cite{PertFirstOrder,CN1}.
Also the damping rate mentioned above in perturbation theory is found to
diverge at the phase-transition, yielding "critical speeding up" instead of the
expected slowing down.

A method which is well known in statistical physics and has become more widely
spread in the particle theory community in recent years is the "Wilsonian
renormalization group" invented specifically to deal with situations involving
divergent correlation lengths
\cite{Wilson}.
The first application of this formalism to field theory is due to Polchinksi
(\cite{Polchinski}, see also 
\cite{Hasenfratz}).
The standard way of adapting the method for a relativistic field theory works
in Euclidean space-time, and there are a number of concrete realizations of
this approach.
These range from the "environmentally friendly" RG (
\cite{ChrisReview} and references therein)
via the "auxiliary mass method"
\cite{Drummondetal,Satoetal}
to the "exact renormalization group" (ERG) of Wetterich
\cite{Christof}.
These methods are straightforwardly adapted to finite temperature calculations
in the Matsubara-formalism.
However, as pointed out above, there are a number of interesting
quantities that 
may not be obtained in a direct way from calculations in the imaginary-time
formalism and it thus is attractive to devise methods for calculations directly
in real time that are also able to deal with critical theories.
One such method is the "thermal renormalization group" (TRG), proposed by
D'Attanasio and Pietroni
\cite{DAP1}
for scalar theories and later also generalized to gauge-theories
\cite{DAP2}.
It was shown in 
\cite{B1} and
\cite{BJ1}
that this method provides a good tool to study the phase-transition of (at
vanishing temperature weakly coupled) scalar theories.
Also a first calculation of the behavior of the damping rate of a scalar
$\varphi^4$-theory close to its critical temperature was performed in 
\cite{Massimo} and critical slowing down was indeed observed.

In the present paper we will be concerned with the formulation of the TRG for
theories with fermionic degrees of freedom.
As an example we will study a simple model displaying a spontaneously broken
chiral symmetry which is restored at high temperatures.
This model involves two flavors of chiral fermions and an $O(4)$-symmetric
scalar sector with at $T\!=\!0$ spontaneously broken symmetry.
A Yukawa-coupling of the fermions to the order-parameter field then breaks the
chiral symmetry in the presence of a nonvanishing VEV.
This modification of the sigma-model was investigated in 
\cite{ChristofDirk}
and is believed to describe the $N_f=2$ chiral phase-transition of QCD
in the limit where the axial $U(1)$-symmetry is completely broken.
Based on universality arguments the theory should undergo a phase-transition
of second order with critical exponents of an $O(4)$-symmetric scalar model.
After the formulation of the TRG we concentrate on the universal critical
behavior.
It turns out that the standard tool in the application of Wilsonian
renormalization group equations, the derivative expansion, does not work in the
fermionic sector, and we discuss the implications of this observation for
calculations in models with chiral symmetry.
A necessary ingredient for fermion-decoupling in the framework
of the TRG is an effective mass-term which
does not break chiral symmetry and is obtained from considering "hard thermal
loop"-contributions
\cite{fermionmass}.
These result in a complicated dispersion relation for fermionic fields and we
discuss how the propagators may be simplified without loosing
the important effects.

The paper is organized as follows:
In the next section we present the formulation of the thermal renormalization
group equation for theories involving fermionic degrees of freedom.
It will be a straightforward extension of the TRG for scalar theories and for
details of the practical application including necessary approximations we
refer the reader to
\cite{DAP1,BJ1}.
In section 3 we introduce the model we will use.
Section 4 contains the discussion of the fermionic dispersion relations and
constitutes an important part of the present work.
In section 5 we then report on the results we find for the universal aspects of
the critical behavior of the model under consideration.
These compare well with results from other approaches and prove our
method to constitute a good starting point for the nonperturbative exploration
of dynamical aspects of the phase-transition.
Section 6 gives a summary and conclusions.

\section{The thermal renormalization group for fermions}

In this section we discuss the formulation of the thermal renormalization group
(TRG) for fermionic degrees of freedom.
The TRG equation is a Wilsonian renormalization group equation in Minkowski
space for the
dependence of an effective action on an external cutoff scale $\Lambda$, where
this cutoff is imposed only on the thermal fluctuations
\cite{DAP1}.
This is possible since in the real-time formulation of thermal field theory the
propagators contain two parts, one being identical to the
$(T\!=\!0)$-propagator and 
another one being multiplied by the appropriate thermal distribution. 
We will impose a cutoff on this second part.
A somewhat more extensive discussion of the closed time-path approach to
thermal field theory in connection with Wilsonian renormalization group methods
may be found in 
\cite{BJ1}.
We will here assume that the reader is familiar with thermal field theory in
real time, and only discuss the implementation of the cutoff and the
corresponding RG-equations.

As stated above, the suppression of soft contributions to the path integral is
implemented by modifying the two-point functions. 
In the operator approach, the propagators are obtained as the expectation
values of time-path-ordered products of two field operators at different space-time
points.
Introducing for fermionic fields the Fourier decomposition
\begin{eqnarray}
  \hat \psi(x) &=& \sum_{\pm s} \int \frac{d^3 k}{(2 \pi)^3 2 \omega_k}
  \left[ b(k,s) u(k,s) e^{(-i k x)} + d^{\dagger} (k,s) v(k,s) e^{(i k x)}
  \right]
\nonumber \\
  \hat \psi^{\dagger} (x) &=& \sum_{\pm s} \int \frac{d^3 k}{(2 \pi)^3 2
  \omega_k}
  \left[ b^{\dagger} (k,s) u^{\dagger} (k,s) e^{(i k x)} + d (k,s)
  v^{\dagger} (k,s) e^{(-i k x)} \right] 
\label{2.1}
\end{eqnarray}
the propagator depends on the expectation values of products of $b$,
$b^\dagger$, $d$, and $d^\dagger$.
In thermal equilibrium at a given temperature $T=\beta^{-1}$ one has
\begin{eqnarray}
  \left< b^+(k,s) b(k',s') \right>_{\beta} &=& (2 \pi)^3 2 
  \omega_k \tilde{N}(\omega_k) 
  \delta_{s s'}  \delta(\vec k - \vec k')
\nonumber \\
  \left< d^+(k,s) d(k',s') \right>_{\beta} &=& (2 \pi)^3 2 \omega_k
  \tilde{N}(\omega_k) \delta_{s s'} 
  \delta(\vec k - \vec k') 
\nonumber \\
  \left< b(k,s) b^+(k',s') \right>_{\beta} &=& (2 \pi)^3 2 \omega_k
  \left( 1 - \tilde{N}(\omega_k) 
  \right) \delta_{s s'} \delta(\vec k - \vec k') 
\nonumber \\
  \left< d(k,s) d^+(k',s') \right>_{\beta} &=& (2 \pi)^3 2 \omega_k
  \left( 1 - \tilde{N}(\omega_k) 
  \right) \delta_{s s'} \delta(\vec k - \vec k') 
\label{2.2}
\end{eqnarray}
where $\tilde{N}(\omega)$ is the Fermi-Dirac distribution,
\bea
\tilde{N}(\omega) = \frac{1}{e^{\omega/T}+1} \; .
\label{2.3}
\eea
The above expectation values differ from their bosonic counterparts only by the
appearance of $\tilde{N}(\omega)$ instead of the Bose-distribution
$N(\omega)$ and by the minus-sign in the last two equations, this
being due to the anticommuting nature of the fermion fields.
The introduction of the cutoff-scale $\Lambda$ now proceeds in the same way as
in the case of scalar fields
\cite{DAP1}, namely by replacing
\bea
\tilde{N}(\omega_k) \rightarrow \tilde{N}(\omega_k ; \Lambda)
\label{2.4}
\eea
with
\bea
\tilde{N}(\omega_k ; \Lambda) \rightarrow \left\{ \begin{array}{cl}
 \tilde{N}(\omega_k) \quad  & {\mbox{for }}|\vec{k}|\gg \Lambda \\
          0          \quad  & {\mbox{for }}|\vec{k}|\ll \Lambda 
\end{array} \right. \; .
\label{2.5}
\eea
This implies that the "hard" modes are in equilibrium at temperature $T$
whereas "soft" thermal fluctuations are suppressed.
We will write
\bea
\tilde{N}(\omega_k; \Lambda) = \Theta(|\vec{k}|,\Lambda) \tilde{N}(\omega_k)
\label{2.6}
\eea
and in practical applications always use a sharp cutoff with
$\Theta(|\vec{k}|,\Lambda) = \theta(|\vec{k}|-\Lambda)$.
The cutoff thermal propagator then reads ($\alpha$ and $\beta$ are spinor
indices, $T_c$ denotes time-ordering on the contour $C$)
\begin{equation}
  iS_{c,\Lambda} (x,x')_{\alpha \beta} = \left< T_c \psi_{\alpha} (x) \bar
  \psi_{\beta} (x') \right>^{\Lambda}_{\beta} = \theta_c (t - t')
  S_{c,\Lambda}^> (x,x')_{\alpha \beta} + 
  \theta_c (t' - t) S_{c,\Lambda}^< (x,x')_{\alpha \beta}
\label{2.7}
\end{equation}
with
\begin{equation}
  S_{c,\Lambda}^> (x,x')_{\alpha \beta} = \left< \psi_{\alpha} (x) \bar
  \psi_{\beta} (x') \right>_{\beta}^\Lambda \;,\;\;\;
  S_{c,\Lambda}^< (x,x')_{\alpha \beta} = - \left< \bar \psi_{\beta} (x')
  \psi_{\alpha} (x') \right>_{\beta}^\Lambda \; .
\label{2.8}
\end{equation}
Inserting \eqref{2.1} and \eqref{2.2} with the modification \eqref{2.4} thus gives
\begin{eqnarray}
  iS_{c,\Lambda} (x - x')_{\alpha \beta} &=& \intg4k \; 2 \pi \delta ( k^2
  - m_0^2 ) ( \slash k + m_0 )_{\alpha \beta} e^{-i k (x -x')} 
\nonumber \\
  && \left[
  \theta_c (t - t') \theta ( k_0 ) + \theta_c ( t' - t ) \theta ( - k_0 )
  - \tilde{N} ( |k_0| ) \Theta( |\vec k|, \Lambda ) \right] \; .
\label{2.9}
\end{eqnarray}
Now we may adopt the matrix-notation which has become common in real-time
thermal field theory (see 
\cite{LeBellac})
and write the fermion propagator as
\begin{equation}
  S_{\Lambda} (k) = \left( 
  \begin{array}{cc} 
   \Delta_0^f & (\Delta_0^f - {\Delta_0^f}^*) \theta(- k_0)
  \\
   (\Delta_0^f - {\Delta_0^f}^*) \theta(k_0) & - {\Delta_0^f}^*
  \end{array} \right) - (\Delta_0^f - {\Delta_0^f}^*) \tilde{N}
    ( |k_0| ) \theta ( |\vec k| - \Lambda ) \left( 
  \begin{array}{cc} 1 & 1 \\ 1 & 1 \end{array} \right) 
\label{2.10}
\end{equation}
with
\begin{equation}
  \Delta_0^f = \left[ \slash k - m_0 + i \epsilon \right]^{-1} \; ,
  \;\;\;
  {\Delta_0^f}^* = \left[ \slash k - m_0 - i \epsilon \right]^{-1} \; .
\label{2.11}
\end{equation}
where we have already switched to a $\theta$-function cutoff.
Together with an analogous expression for the modified scalar propagator 
\cite{DAP1}
we then define the scale dependent partition function
$Z_\Lambda[J,\eta,\bar{\eta}]$ as
\bea
Z_\Lambda[J,\eta,\bar{\eta}] &=& \int \delta \phi \delta \psi \delta
\bar \psi \exp \Biggl \lbrace 
i \Big( \frac{1}{2} \phi \cdot D_\Lambda^{-1} \cdot \phi + \bar{\psi} \cdot
S_\Lambda^{-1} \cdot \psi + \nonumber \\
&& \mbox{\hspace{2.6cm}} + J \cdot \phi + \bar{\eta} \cdot \psi +
\bar{\psi}
\cdot \eta + S_{\mathrm{int}}[\phi,\psi,\bar{\psi}] \Big) \Biggr
\rbrace \; .
\label{2.12}
\eea
From the way the cutoff was introduced, it is obvious that
\bea
Z_{\Lambda \rightarrow \infty}[J,\eta,\bar{\eta}] =
Z^{(T=0)}[J,\eta,\bar{\eta}]
\label{2.14}
\eea
and that in the opposite limit $\Lambda \rightarrow 0$ the usual generating
functional for the theory at temperature $T$ is obtained.
Finally, since all scale-dependence of $Z_\Lambda$ is explicit and contained in
the terms bilinear in the fields, it is straightforward to obtain an expression
for the $\Lambda$-dependence of $Z_\Lambda$.

As in the Euclidean formulation it is however more convenient to work with a
slightly modified version of the effective action, the generating functional
for the 1PI-Green functions. This is obtained from the Legendre transform of
the logarithm of $Z_\Lambda$ according to
\bea
\Gamma_\Lambda[\varphi,\psi,\bar{\psi}] &=& -i \ln Z_\Lambda - J\cdot \varphi -
\bar{\eta} \cdot \psi - \bar{\psi} \cdot \eta - \frac{1}{2} \varphi \cdot
D_\Lambda^{-1} \cdot \varphi - \bar{\psi} \cdot S_\Lambda^{-1} \cdot
\psi \; .
\label{2.15}
\eea
Here the fields are now the "classical" fields as $\varphi =
\frac{\delta(-i \ln Z_\Lambda)}{\delta J}$ etc.
The sources on the right hand side are to be regarded as functionals of the
fields, given by the inversion of the above expressions, and are thus
scale-dependent.
The RG-equation governing the scale-dependence of $\Gamma_\Lambda$ is now
easily obtained and one finds
\begin{equation}
  \Lambda \partial_{\Lambda} \Gamma_{\Lambda} [ \Phi ]
  = \frac{i}{2} \mbox{STr} \left\{ (\Lambda \partial_{\Lambda} {\cal
  D}_{\Lambda}^{-1} ) \left( \frac{ \delta^2 \Gamma_{\Lambda} }{
  \delta \bar \Phi \delta \Phi } + {\cal D}_{\Lambda}^{-1} \right)^{-1}
  \right\} 
 \label{master_eq}
\end{equation}
which is a straightforward extension of the scalar TRG-equations 
\cite{DAP1}.
$\Phi$ denotes a superfield
\bea
\Phi = (\phi,\psi,\bar{\psi}) \quad ; \quad \bar{\Phi} =
(\phi,-\bar{\psi},\psi) 
\label{2.13}
\eea
and the supertrace takes into account the anticommuting nature of the
Grassmann-fields. Note that we have suppressed all indices: The fields
carry a spacetime-index, a thermal index, possible internal indices
and (for the fermions) Dirac indices. All these are traced in
\eqref{master_eq}. Also by our conventions the bare inverse
propagator-matrix reads 
\bea
  {\cal D}_{\Lambda}^{-1} (k)  = \left( 
  \begin{array}{ccc} D_{\Lambda}^{-1}(k) & & \\ 
                      & S_{\Lambda}^{-1}(k) & \\
                      & & (S_{\Lambda}^{-1}(k))^T 
  \end{array} \right) 
\label{2.17}
\eea
and we have
\bea
\frac{\delta^2 \Gamma_{\Lambda}}{\delta \bar \Phi \delta \Phi}
\equiv \Gamma_{\Lambda}^{(2)}
  = \left( 
  \begin{array}{ccc}
   \frac{\delta^2 \Gamma_{\Lambda}}{\delta \phi \delta \phi} &
   \frac{\delta^2 \Gamma_{\Lambda}}{\delta \phi \delta \psi} &
   \frac{\delta^2 \Gamma_{\Lambda}}{\delta \phi \delta \bar \psi} 
  \\
   - \frac{\delta^2 \Gamma_{\Lambda}}{\delta \bar \psi \delta \phi} &
   - \frac{\delta^2 \Gamma_{\Lambda}}{\delta \bar \psi \delta \psi} &
   - \frac{\delta^2 \Gamma_{\Lambda}}{\delta \bar \psi \delta \bar \psi} 
  \\
   \frac{\delta^2 \Gamma_{\Lambda}}{\delta \psi \delta \phi} &
   \frac{\delta^2 \Gamma_{\Lambda}}{\delta \psi \delta \psi} &
   \frac{\delta^2 \Gamma_{\Lambda}}{\delta \psi \delta \bar \psi} 
  \end{array} \right) \; .
\label{2.18}
\eea
The general properties of the TRG-equation are extensively discussed in 
\cite{DAP1,BJ1} and we will here only make a few remarks in order to
make the present work self-contained.
Let us first note that no approximations have been made at this point. 
The TRG-equation \eqref{master_eq} is an exact relation on the same level as
for example the Schwinger-Dyson equations.
However, it is of course impossible to give a general solution to this
functional differential equation -- note that the matrix
$\left(\Gamma_\Lambda^{(2)} + {\cal D}_{\Lambda}^{-1} \right)^{-1}$ is the full
propagator.
The equation is formally a one-loop equation and we may introduce the
derivative $\tilde{\partial}_\Lambda$ which is defined to be a derivative
acting only on the explicit scale-dependence in the bare propagator.
Using this derivative we have
\bea
\Lambda \partial_\Lambda \Gamma_\Lambda[\Phi] = \frac{i}{2} \mbox{STr} \Lambda
\tilde{\partial}_\Lambda \ln \left(\Gamma_\Lambda^{(2)} + {\cal
D}_{\Lambda}^{-1} \right)
\label{2.19}
\eea
and the one-loop character is apparent.
We may thus give a simple recipe for calculations in the framework of the
thermal renormalization group: To derive the flow-equation of a specific
coupling, write down the one-loop diagrams that would contribute to this
coupling, taking into account all vertices allowed by the symmetries of
the theory under consideration.
The flow-equation is then obtained by considering only those parts of the
diagram which involve thermal distribution functions (since the
$\Lambda$-dependence only comes with those in the TRG-approach), replacing the
distribution functions according to \eqref{2.4} and similarly for the
Bose-distribution, taking the derivative with respect to the explicit
scale-dependence and then performing the loop-integral.
It is obvious that this integral will be finite. 
The UV-behavior of the thermal parts of any Feynman-diagram is never
problematic since any UV-divergences of the full theory at finite temperature
reside in the ($T\!=\!0$)-parts of the diagrams.
The behavior in the infrared, which might be problematic in perturbation
theory, is regulated by the introduction of the external scale (see
\eqref{2.5}).
The formal exactness of this approach is due to the use of full
propagators and vertices.

In order to simplify the calculations somewhat we may make use of an
observation made in 
\cite{NS},
namely the fact that one may get rid of the fields on the lower part of the
CTP-contour (the "thermal ghost fields") if one is interested in physical
observables.
To this end one may introduce a functional $\bar{\Gamma}$ which depends on one
(super)field $\Phi$ (without thermal indices) and is defined through
\bea
\bar{\Gamma}[\Phi] = \left. \Gamma[\Phi_1,\Phi_2[\Phi_1]]
\right|_{\Phi_1=\Phi}
\label{2.20}
\eea
Here $\Phi_2$ is the solution to its equation of motion for arbitrary $\Phi_1$,
i.e. we have
\bea
\left. \frac{\delta \Gamma[\Phi_1,\Phi_2]}{\delta \Phi_2} \right|_{\Phi_2 =
\Phi_2[\Phi_1]} = 0
\label{2.21}
\eea
For space-time independent fields one has $\Phi_2[\Phi_1] = \Phi_1$
and \eqref{2.20} is up to a constant equivalent to
\bea
\frac{\delta \bar{\Gamma}[\Phi]}{\delta \Phi} = \left. \frac{\delta
\Gamma[\Phi_1,\Phi_2]}{\delta \Phi_1}
\right|_{\Phi_1=\Phi_2=\Phi=\mbox{const.}} \; . 
\label{2.22}
\eea
This is the definition that we will use below.
The TRG-equation for $\bar{\Gamma}_\Lambda^{(1)}$ (which is obtained from
\eqref{2.22} after replacing $\Gamma$ with $\Gamma_\Lambda$) is readily found
from \eqref{master_eq} and \eqref{2.22}.
Some care is required in the presence of fermionic fields, since
derivatives with respect to fermionic fields do not commute with a supermatrix
like the full propagator. 
We have for example
\bea
\frac{\delta}{\delta \varphi} \Gamma_\Lambda^{(2)} &=& \frac{\delta
\Gamma_\Lambda^{(2)}}{\delta \varphi} + \Gamma_\Lambda^{(2)}
\frac{\delta}{\delta \varphi} \nonumber \\
\frac{\delta}{\delta \psi} \Gamma_\Lambda^{(2)} &=& \frac{\delta
\Gamma_\Lambda^{(2)}}{\delta \psi} + M \Gamma_\Lambda^{(2)} M
\frac{\delta}{\delta \psi} 
\label{2.23}
\eea
where $\varphi$ ($\psi$) is a bosonic (fermionic) field and $M =
\mbox{diag}(1,-1,-1)$ (in the present case).
The supertrace is cyclic for supermatrices.
Note also that the fermionic derivative of a supermatrix (e.g. $\frac{\delta
\Gamma_\Lambda^{(2)}}{\delta \psi}$) is not a supermatrix.

In the following section we will first introduce
the model that we will be interested in and then turn to a discussion of the approximations we have to make in order to convert the
functional differential equation \eqref{master_eq} into a solvable system of
(partial) differential equations.

\section{The chiral quark-meson model}

We will be interested in a simple model involving only scalar and fermionic
fields with scalar self-coupling as well as Yukawa-couplings.
It is well accepted that the chiral phase-transition of QCD in the
limit of two massless flavors and an explicitly broken axial $U(1)$ should be
of second order, with $O(4)$-model universal behavior.
This could in principle be modeled by a simple $O(4)$-symmetric scalar theory
with (at $T\!=\!0$) spontaneously broken symmetry.
Recently, Jungnickel and Wetterich have pointed out that a modification of the
sigma-model might actually be quantitatively connected to QCD by a
combination of analytical and numerical calculations
\cite{ChristofDirk}.
This model, the so called "chiral quark-meson model", includes (in the limit of
strongly broken $U(1)_A$) the usual sigma-model fields, which are however
coupled to $N_c$ dublets of fermionic fields via a Yukawa-coupling.
As was pointed out above, perturbation theory is not applicable even for small
$(T\!=\!0)$-couplings at this phase-transition.
On the other hand, the Wilsonian RG was introduced to deal with such
situations.
The chiral quark-meson model is thus, apart from being interesting
phenomenologically, an ideal testing ground for our method.

We shall not discuss in detail here how the model may be connected to QCD and
refer the reader to 
\cite{ChristofDirk}
for details.
We will actually in the present work not even try to make quantitative contact
with this effective description by choosing the values of the
$(T\!=\!0)$-couplings accordingly -- this could in principle be done
and we briefly comment on this below.
For the main part of this paper we shall discuss the universal physics obtained
for this model in different approximations. 
These are independent of the initial values of the couplings and thus also
apply to the $N_f = 2$ chiral phase-transition.
Nonuniversal quantities will only be considered for small $(T\!=\!0)$-couplings in
the present work.

The initial condition that we will use for the effective action
$\bar{\Gamma}_\Lambda$ as $\Lambda \rightarrow \infty$, i.e. the effective
action of the $(T\!=\!0)$-theory is (remember that the free two-point functions are subtracted in $\Gamma_\Lambda$, c.f. \eqref{2.15})
\bea
\bar{\Gamma}_{\Lambda\rightarrow \infty} = \int d^4 x \left[ -
\frac{g_{(T=0)}}{2} \rho^2 - \bar{h}_{(T=0)}\bar \psi \left(
\frac{\varphi+\varphi^{\dagger}}{2}  + \gamma_5\frac{\varphi-\varphi^{\dagger}}{2}
\right) \psi \right]
\label{3.1}
\eea
It involves a fermion flavor dublet as well as a scalar multiplet containing
four real degrees of freedom,
\bea
\psi=(\psi^1,\psi^2) \qquad,\qquad
\varphi=\frac{1}{2}(\sigma+i\vec{\pi}\vec{\tau})
\label{3.2}
\eea
Furthermore we use the $O(4)$-symmetry to write
\bea
\rho=\mbox{Tr}(\varphi^{\dagger}\varphi)=\frac{1}{2}(\sigma^2+\vec{\pi}^2)
\label{3.3}
\eea
The fermions come with an additional color index and closed fermion-loops
correspondingly yield a factor of $N_c$.

The effective action given in \eqref{3.1} is of course only a crude
approximation to a full effective action of a $(T\!=\!0)$-theory -- one could
e.g.~include perturbative contributions. 
We will nevertheless use this simple form and report nonuniversal results only
for small values of the initial couplings $g_{(T=0)}$ and
$\bar{h}_{(T=0)}$, such 
that the higher order contributions should be small.
The reason for doing so is that the universal behavior is unaffected by these
approximations and that on the other hand the nonuniversal quantities would be
most interesting using the $(T\!=\!0)$-parameters that correspond to
the realistic 
case where observables calculated from the model \eqref{3.1} match those found
in nature.
A discussion of this matching and the resulting values for the
$(T\!=\!0)$-couplings 
as obtained from Euclidean versions of Wilsonian renormalization group
equations may be found in 
\cite{ChristofDirk}.
One finds for example that the Yukawa-coupling is $\bar{h}^2 \sim
{\mathcal{O}}(10)$ and thus by no means small. 
If we were interested in nonuniversal features of the chiral phase-transition
as obtained from the quark-meson model we would have to address the question of
initial values carefully.
Studies using the Matsubara-formulation of thermal field theory in connection
with the average effective action yield reasonable predictions for e.g.~the
critical temperature
\cite{JNC2}.

Within our approximations the effective action \eqref{3.1} thus constitutes the boundary value for the
solution of the TRG equation \eqref{master_eq} (combined with the definition of
$\bar{\Gamma}_\Lambda$, \eqref{2.20}).
We now have to discuss the approximations we will use in order to bring this
equation into a solvable form.
The simplest approximation one may think of is to assume that the effective
action at finite temperature always has the same form as given in \eqref{3.1}
and that only the couplings -- in this case the scalar massterm, the quartic
coupling and the Yukawa-coupling -- depend on the scale $\Lambda$.
One then obtains the flow-equations of those couplings by plugging
$\bar{\Gamma}_\Lambda$ from \eqref{3.1} into \eqref{master_eq} and matching
coefficients on both sides.
This simple approximation will give completely misleading results, for reasons
to be discussed in the next section.
In the absence of fermions (for $\bar{h}=0$) this approximation would give a
second order phase-transition, however the values for the universal critical
exponents will differ strongly from known results.
The situation can be considerably improved if one allows for a general form of the
effective action for a constant scalar field $\rho(x) = \rho$
\cite{B1,BJ1}. 
The effective action for constant field is connected to the effective potential
through
\bea
\bar{\Gamma}_\Lambda[\varphi=\sqrt{2 \rho}, \psi =
\bar{\psi} = 0] = \left( m^2 \rho - U(\rho) \right) \int d^4 x
\label{3.4}
\eea
for a constant configuration $\rho$ with the bare mass $m$. 
The effective potential corresponds to the free energy for constant fields and
it approaches a universal form at a second order phase-transition.
For the results for this universal form in scalar $O(N)$-models, using the
approximation of a standard kinetic term (as in \eqref{3.1}) and a general
effective potential, see 
\cite{BJ1}.
The critical exponents one finds in this case are in much better agreement with
the known values than in the case of a simple polynomial approximation for $U$
\cite{DAP1}.

Neglecting corrections to the kinetic term may be motivated by arguing in the
framework of an expansion of the effective action in powers of derivatives.
In this expansion, the effective potential is obviously the lowest order term.
The leading term beyond this approximation then is obtained from the scalar
wave-function renormalization, i.e. using the approximation
\begin{equation}\label{gammaqueron}
\bar{\Gamma}_{\Lambda}[\varphi,\psi,\bar{\psi}] = \int d^4 x \left[ (Z_{\Lambda}-1)\mbox{Tr} \left( \partial^{\mu} \varphi^{\dagger} \partial_{\mu}
\varphi \right) +m^2\rho - U_{\Lambda}(\rho) \right] + {\mathrm{Fermion-contributions}}
\end{equation}
Here the wave-function renormalization $Z_\Lambda$, which could in principle be
a function of the field, is evaluated at the minimum of the effective potential.
This will be the approximation we use in the scalar sector of the model.
It is widely used also in Euclidean versions of the Wilsonian renormalization
group, see e.g. 
\cite{JNC2}.
In particular we will neglect all imaginary contributions to the two-point
functions.
These would only indirectly affect the scale dependence of the effective
potential through a contribution to the anomalous dimension.
Since the anomalous dimension will turn out to be very small in all situations
that we will be interested in, we believe that dropping imaginary contributions
results in negligible errors in quantities like the critical temperature and
the critical exponents (with the possible exception of $\eta$, see below).
On the other hand the imaginary parts are interesting objects by themselves and
we plan to study them in future work.

Let us now turn to the sector involving fermionic fields.
Here the situation is more complicated. 
The reason for the complication is that in contrast to the bosonic sector,
where the leading finite temperature effects show up in a local
contribution to the mass $\propto T$ in lowest order derivative
expansion, there is no local massterm generated for the fermions.
We will come back to this issue in section 4 and for now only consider
the interaction term. 
Here one obtains to lowest order in the derivative expansion a
general function of the fields (which should respect the $SU(2) \times SU(2)
\times U(1)$-symmetry of \eqref{3.1}). 
However, one expects the fermions to decouple from the infrared physics and
universal quantities should be independent of the approximations in the
fermionic sector. 
The dependence of nonuniversal quantities as the critical temperature should in
principle be calculable in perturbation theory and we may, as long as we
restrict ourselves to small values of $\bar{h}_{(T=0)}$ and $g_{(T=0)}$, always
argue that any induced coupling involving fermions are of higher order
in the $(T\!=\!0)$-couplings and thus negligible compared to the
Yukawa-coupling. 
We will thus approximate the interaction between fermions and bosons by the
original form as given in \eqref{3.1} for all values of $\Lambda$. 
In particular, we always evaluate the Yukawa-coupling at $\rho=0$ and at
vanishing energies and momenta.

Let us then discuss as an example the flow-equation for the $\rho$-derivative
of the effective potential.
This is obtained from the flow-equation for the $\varphi$-derivative of
$\bar{\Gamma}_\Lambda$ at constant $\varphi$ and vanishing $\bar{\psi}$ and
$\psi$:
\bea
\Lambda \partial_\Lambda \left( \varphi U'(\rho) \right) &=&
\left. \Lambda \partial_\Lambda \frac{\delta \bar{\Gamma}_\Lambda}{\delta
\varphi} \right|_{\varphi={\mathrm{const.}},\bar{\psi}=\psi=0} \nonumber \\
&=& \left. \frac{i}{2} \mbox{STr} \left[ {\cal K}_{\Lambda}
  \frac{\delta^3 \Gamma_{\Lambda}}{\delta \varphi \delta \bar
  \Phi \delta \Phi} \right] \right|_{\varphi={\mathrm{const.}},\bar{\psi}=\psi=0}
\label{3.5}
\eea
Here we have introduced the kernel
\bea
  {\cal K}_{\Lambda} = - \left( \frac{\delta^2 \Gamma_{\Lambda}}
  {\delta \bar \Phi \delta \Phi} 
  + {\cal D}_{\Lambda}^{-1} \right)^{-1} ( \Lambda \partial_{\Lambda}
  {\cal D}_{\Lambda}^{-1} ) \left( \frac{\delta^2
  \Gamma_{\Lambda}}{\delta \bar \Phi \delta \Phi} 
  + {\cal D}_{\Lambda}^{-1} \right)^{-1} = \left( 
  \begin{array}{ccc} K_{\Lambda}^b & & \\
                      & K_{\Lambda}^f & \\
                      & & (K_{\Lambda}^f)^T 
  \end{array} \right)
\label{3.6}
\eea
The scalar kernel may be rewritten as
\cite{DAP1}
\bea
  K^b_{\Lambda} &=& -\Lambda \delta(|\vec k| - \Lambda)
  \Delta_{\Lambda}^b (\Delta_{\Lambda}^b)^* \frac{\Delta_0^b -
  (\Delta_0^b)^*}{\Delta_0^b (\Delta_0^b)^*} N(|k_0|) \left( \begin{array}{cc} 1 & 1 \\ 1 & 1 \end{array}
  \right) 
\label{kernelscalar}
\eea
The fermion-kernel is obtained from \eqref{kernelscalar} by replacing
the bare and full bosonic propagator $\Delta_0^b$ and
$\Delta_\Lambda^b$ by the corresponding fermionic ones and
substituting the Bose-distribution $N(\omega)$ by the minus
Fermi-Dirac one. Note that we have already inserted a sharp cutoff for
\eqref{kernelscalar}. 
Due to the thermal matrix-structure of the kernel and the fact that
\cite{NS}
\bea
\frac{\delta^3 \bar{\Gamma}_\Lambda}{\delta \Phi \delta \bar\Phi \delta \Phi} =
\left. \sum_{i,j=1,2} \frac{\delta^3}{\delta \Phi_i \delta \bar{\Phi}_j \delta \Phi_1}
\Gamma_\Lambda \right|_{\Phi_2 = \Phi_1 = \Phi}
\label{3.8}
\eea
the trace over the thermal indices in \eqref{3.5} allows us to rewrite the right hand side in terms of $\bar{\Gamma}_\Lambda$.
Furthermore for real self-energies we may write
\bea
\Delta_{\Lambda}^b (\Delta_{\Lambda}^b)^* \frac{\Delta_0^b -
  (\Delta_0^b)^*}{\Delta_0^b (\Delta_0^b)^*} = 2 \pi i \rho^b_\Lambda(q)
\label{3.10}
\eea
with the spectral density $\rho_\Lambda$ and similarly for the fermions.
The spectral densities may be obtained from the ansatz for the effective action
$\bar{\Gamma}$. 
In our case, using the approximations discussed above for the scalar fields and
for the moment dropping any momentum-dependent contribution other than the
standard kinetic term for the fermions, we find 
\bea
2 \pi i \rho^b_\Lambda(q) &=& \frac{1}{Z_\Lambda q^2 - M^2 + i \epsilon} 
 - \frac{1}{Z_\Lambda q^2 - M^2 - i \epsilon} \nonumber\\
2 \pi i \rho^f_\Lambda(q) &=& \frac{\slash q + m_{\chi SB}}{q^2 - m_{\chi
SB}^2 + i \epsilon}-\frac{\slash q + m_{\chi SB}}{q^2 - m_{\chi SB}^2
- i \epsilon}  
\label{3.11}
\eea
where $m_{\chi SB}^2 = \bar{h}^2 \rho / 2$ and $M^2$ are the
eigenvalues of the matrix $\frac{\partial^2 U}{\partial \varphi
\partial \varphi}$.
For $\epsilon \rightarrow 0$, the spectral densities yield $\delta$-functions
which allow us to perform the $q_0$-integration implicit in the trace in
\eqref{3.5}. 
The angular integral is trivial for constant fields.
Finally from the $\Lambda$-derivative of the cutoff, we have another
$\delta$-function (c.f.~\eqref{kernelscalar}) which allows us to perform the
integration with respect to the modulus of the loop-momentum.
The result for the flow-equation of the effective potential is
\bea
\Lambda \partial_\Lambda U'(\rho)\!\!&=&\!\! - \frac{\Lambda^3}{4\pi^2} \left[ \left(
3 U''(\rho) + 2 \rho U^{(3)}(\rho) \right)
\frac{N(\bar{\omega}_\sigma)}{\bar{\omega}_\sigma} \Theta(\bar{\omega}_\sigma^2)+ 3 U''(\rho)
\frac{N(\bar{\omega}_\pi)}{\bar{\omega}_\pi}\Theta(\bar{\omega}_\pi^2) \right] -
N_c \frac{\Lambda^3}{\pi^2} \bar{h}^2
\frac{\tilde{N}(\bar{\omega}_\psi)}{\bar{\omega}_\psi} 
\label{4.1}
\eea
with 
\bea
\bar{\omega}_\sigma^2 &=& Z_\Lambda \Lambda^2 + U'(\rho) + 2 \rho U''(\rho) \nonumber \\
\bar{\omega}_\pi^2 &=& Z_\Lambda \Lambda^2 + U'(\rho) \nonumber \\
\bar{\omega}_\psi^2 &=& \Lambda^2 + \frac{\bar{h}^2 \rho}{2}
\label{3.13}
\eea
The $\Theta$-functions arise from the bosonic spectral density
\cite{DAP1}
and will be understood in the following.
The structure of \eqref{4.1} is again easy to understand from the one-loop
character of the flow-equation: the first contribution is from the radial mode
which is massive at the minimum of $U(\rho)$, the second contribution is from
the 3 Goldstone-bosons which are massless at the minimum, and the last
contribution is from the $2 N_c$ fermionic fields which are degenerate.

Similar flow-equations may be obtained for the Yukawa-coupling and the scalar
wave-function renormalization constant.
Before we come to an investigation of the critical behavior of the model using
these flow-equations in section 5, we will in the next section turn to the
question of approximations to the fermionic two-point function.

\section{Fermion decoupling, universality and hard thermal loops}

In this section we will discuss how the fermions decouple from the critical
behavior of the quark-meson model, leading to a second order phase-transition
with $O(4)$-model critical behavior.
We will argue that the mechanism of fermion decoupling at a second order
phase-transition is different from the mechanism behind fermion-decoupling at
high temperatures. 

Let us start with discussing the high-temperature limit of our model. 
We will work for the moment with the simplest approximation to the full flow
equation for the effective action and only take into account the effective
potential and the Yukawa-coupling and set $Z_{\Lambda}=1$.
The flow-equation for the Yukawa-coupling reads within these
approximations 
\bea
\Lambda \partial_\Lambda \bar{h} &=& \frac{\Lambda^3}{4 \pi^2} \bar{h}^3
\frac{1}{U'(0)} \left[ \frac{N(\bar{\omega}_0)}{\bar{\omega}_0} +
\frac{\tilde{N}(\Lambda)}{\Lambda} \right]
\label{4.2}
\eea
where 
\bea
\bar{\omega}^2_0 &=& \Lambda^2 + U'(0) \; .
\label{4.3}
\eea
Considering now the limit of $T$ large compared to all scales in the theory we
may use the asymptotic expressions 
\bea
N(x) \;
{\lower2.5pt\hbox{$\stackrel{\longrightarrow}{{\scriptscriptstyle{T
\gg x}}}$}} \; \frac{T}{x}  
\label{4.4}
\eea
and
\bea
\tilde{N}(x) \;
{\lower2.5pt\hbox{$\stackrel{\longrightarrow}{{\scriptscriptstyle{T
\gg x}}}$}} \; \frac{1}{2}
\label{4.5}
\eea
to obtain the system of flow-equations (\ref{4.1}) and (\ref{4.2}) in the
high-temperature limit
\bea
\left. \Lambda \partial_\Lambda U'(\rho) \right|_{\mathrm{high-T}} &=& - \frac{T
\Lambda^3}{4 \pi^2} \left[ \frac{3 U''(\rho) + 2 \rho U^{(3)}(\rho)}{\Lambda^2
+ U'(\rho) + 2 \rho U''(\rho)} + 3 \frac{U''(\rho)}{\Lambda^2 + U'(\rho)}
\right] + ... \nonumber \\
\left. \Lambda \partial_\Lambda \bar{h} \right|_{\mathrm{high-T}} &=&
\frac{T \Lambda^3}{4\pi^2} \bar{h}^3
\frac{1}{U'(0) \left( \Lambda^2 + U'(0) \right)} + ... \; .
\label{4.6}
\eea
The contribution of the fermions to the flow of the effective
potential thus vanishes in the high-temperature limit, due to the different
behavior of the distribution functions \eqref{4.4} and \eqref{4.5}.
This is what is referred to as "fermion decoupling" and the TRG thus
reproduces the fact that the dynamics of the soft modes in the high-temperature
limit is independent of the fermionic content of the theory.

Let us now turn to the question about fermion-decoupling at a second order
phase-transition. 
One is interested in this case in a (three-dimensional) scaling
solution for the free energy (i.e. the effective potential).
In order to investigate this scaling solution it is useful to introduce
rescaled variables according to 
\cite{B1}
\bea
\kappa = \frac{\rho}{\Lambda T} \quad , \quad u(\kappa) = \frac{U(\rho)}{\Lambda^3 T} \quad ,
\quad u^{(n)} = \Lambda^{n-3} T^{n-1} U^{(n)}
\label{4.7}
\eea
At this point we do not rescale the Yukawa-coupling, since if we assume the
fermions to decouple there is no reason why this coupling should approach a
scaling solution.
Introducing furthermore the dimensionless flow-parameter 
\bea
\lambda = \frac{\Lambda}{T}
\label{4.8}
\eea
we obtain for the flow of $u'(\kappa)$ from \eqref{4.1}
\bea
\lambda \partial_\lambda u'(\kappa) = -2 u'(\kappa) + \kappa u''(\kappa) -
\frac{\lambda}{4\pi^2} \left[ \left( 3 u'' + 2 \kappa u^{(3)} \right)
\frac{n(\lambda \omega_\sigma)}{\omega_\sigma} + 3 u'' \frac{n(\lambda
\omega_\pi)}{\omega_\pi} \right] -
N_c \frac{1}{4\pi^2} \bar{h}^2 \frac{\tilde{n}(\lambda
\omega_\psi)}{\omega_\psi}
\label{4.9}
\eea
with the dimensionless quantities
\bea
\omega_\sigma^2 &=& 1 + u'(\kappa) + 2 \kappa u''(\kappa) \nonumber \\
\omega_\pi^2 &=& 1 + u'(\kappa) \nonumber \\
\omega_\psi^2 &=& 1 + \frac{\bar{h}^2 \kappa}{2 \lambda} \nonumber \\
\label{4.10}
\eea
and $n(x) = \left( \exp(x) - 1 \right)^{-1}$, $\tilde{n}(x) = \left( \exp(x) +
1 \right)^{-1}$.
The flow-equation for the Yukawa-coupling is rewritten accordingly as
\bea
\lambda \partial_\lambda \bar{h} = \bar{h}^3 \frac{1}{4\pi^2} \frac{1}{u'(0)}
\left[ \frac{n(\lambda \omega_0)}{\omega_0} + \tilde{n}(1) \right]
\label{4.11}
\eea
where $\omega_0 = 1 + u'(0)$.
We will see that the problems with decoupling within the present approximation
concern the region of small $\kappa$, thus to simplify the discussion and to
make our point clear we expand the effective potential in
powers of $\kappa$. 
Introducing couplings according to 
\bea
u(\kappa) = \sum_n \frac{u_n}{n !} \kappa^n
\label{4.12}
\eea
we obtain flow-equations for the $u_n$ from \eqref{4.9}. 
Let us for example consider the scale dependence of $u_1$, which is directly
obtained from \eqref{4.9} by setting $\kappa=0$. 
It involves the scalar  massterm $u_1$ itself as well as the scalar four-point
coupling $u_2$ (remember $\rho = \frac{\varphi^2}{2}$) and the Yukawa-coupling
$\bar{h}$. 
A scaling solution for the scalar couplings $u_n$ may arise only in the
three-dimensional limit, i.e. for $\lambda \rightarrow 0$. 
In order to identify the corresponding fixed-point values of the couplings we
thus expand the flow-equations.
In the case of $u_1$ we have 
\bea
\lambda \partial_\lambda u_1 = -2 u_1 - \frac{3}{2\pi^2} \frac{u_2}{1+u_1} -
N_c \frac{1}{8 \pi^2} \bar{h}^2 + {\mathcal{O}}(\lambda)
\label{4.13}
\eea
and the fermionic contributions obviously do not decouple -- a possible
fixed-point value would depend on $\bar{h}$ and $N_c$, in contradiction with
the assumption of universality!
It is not hard to see what spoils the high-temperature argumentation in the
present case: It is the fact that we should really not compare the behavior of
the distribution-functions alone, but rather the
contribution of couplings times the distribution-functions as $\lambda
\rightarrow 0$. 
Then it is clear that for a scaling solution (i.e. at a second order
phase-transition) while the scalar distribution function diverges for small
$\lambda$ as $\lambda^{-1}$, the four scalar-coupling multiplying it is
renormalized to $0$ with the same power of $\lambda$. 
The product remains finite for $\lambda \rightarrow 0$.
In contrast, the fermion distribution function does not diverge, but on the
other hand the Yukawa-coupling is not driven to $0$ with $\lambda$ -- the
product also remains finite and is thus for all $\lambda$ of the same order as
the scalar contribution.

This is a particular feature of the critical point. For $T \neq T_c$, away
from the second order phase-transition, the scalar coupling is not renormalized
to zero and indeed the product of coupling times distribution function
for large $\frac{T}{\omega}$ dominates the fermionic contribution as shown above.

Also notice that the presence of a fermion mass-term (as for $\kappa \neq 0$)
changes the situation. 
While the behavior of coupling times distribution
function does not change ($\tilde{n}$ still approaches a constant), now the
factor $\omega_\psi^{-1}$ in the fermion-contribution to $\lambda
\partial_\lambda u'(\kappa)$ behaves as $\sqrt{\frac{2 \lambda}{\bar{h}^2
\kappa}}$ and suppresses the fermion-contribution.
This observation will help us to see how universality is saved and
fermion-decoupling works, a question to which we turn now.

To this end, let us turn again to the truncation used to derive \eqref{4.1} and
\eqref{4.2}. 
We have pointed out that we have neglected all terms beyond the effective
potential and the Yukawa-coupling. 
In a purely scalar theory, it is known that using only the effective potential
one already obtains quite reasonable results for the critical behavior (see
e.g.
\cite{BJ1}).
This is due to the fact that for the critical theory one is effectively
using an expansion in the (small) anomalous dimension.
There is however one other important aspect of the thermal effects on scalar
fields which is fully covered by the approximation with only an effective
potential -- the generation of a thermal mass.
One may easily check that within the present approximation one reproduces the
"hard thermal loop" (HTL-) result for the scalar mass term:
Consider the present model for simplicity with $N_c = 0$. 
The flow-equation for the mass then reads from above
\bea
\Lambda \partial_\Lambda U'(\rho) &=& - \frac{\Lambda^3}{4\pi^2} \left[ \left(
3 U''(\rho) + 2 \rho U^{(3)}(\rho) \right)
\frac{N(\bar{\omega}_\sigma)}{\bar{\omega}_\sigma} + 3 U''(\rho)
\frac{N(\bar{\omega}_\pi)}{\bar{\omega}_\pi} \right]  
\label{4.14}
\eea
Adopting a perturbative point of view, neglecting $U^{(3)}(\rho)$ and
replacing in the spirit of the HTL-approximation $\bar{\omega}_\sigma =
\bar{\omega}_\pi = \Lambda$, we obtain
\bea
\partial_\Lambda U'(\rho) = - 6 \frac{\Lambda}{4\pi^2} U''(\rho)
\frac{1}{e^{\Lambda/T}-1} \; .
\label{4.15}
\eea
Remembering that the boundary condition for the flow is the
$(T\!=\!0)$-effective 
action and neglecting -- again to leading order in perturbation theory -- the
scale-dependence of $U''(\rho)$ we easily solve the flow-equation \eqref{4.15}
to find the temperature-dependent massterm to be given by
\bea
U'(\rho,T) = U'(\rho,T=0)+\frac{U''(\rho)}{4} T^2
\label{4.16}
\eea
reproducing the leading order perturbative result.
The positive contribution $\propto T^2$ of course is also the reason for the
restoration of the spontaneously broken symmetry.
The reason why this leading order high-temperature result is reproduced in the
simple truncation with only the effective potential is that it is obtained
as a local contribution to the effective action. 
Of course, this local contribution is fully resummed if we solve the
flow-equation \eqref{4.1} without further approximations.

In contrast to this simple behavior of the leading thermal contributions in
the scalar case, for the fermionic fields there is no local massterm generated
by hard thermal loops.
It is however well known that the leading order corrections to the fermion
propagator lead to modified dispersion relations which for small momenta
resemble the ones of a massive particle without breaking chiral invariance for
$\kappa=0$
\cite{LeBellac}.
If we neglect thermal contributions to the momentum dependence of the fermion
two-point function, or in fact for any truncation that assumes analyticity for
the momentum dependence of the fermion two-point function, we will miss this
important finite temperature contribution.
The effect is numerically small in the presence of a large enough chiral
symmetry breaking mass, but in the situation discussed above with $\kappa=0$,
this effective mass-term takes care of the decoupling of the fermionic
fluctuations, as we will argue in the following.

Let us first briefly review the structure of the leading contribution to
the fermion self-energy for large $T$ in perturbation theory
\cite{LeBellac}.
The contributing diagram is given in figure 1. 
\begin{figure}
\begin{minipage}{16.5cm}
\begin{center}
\begin{picture}(240,120)(0,0)
\ArrowLine(0,60)(80,60)
\Vertex(80,60) {1.5}
\ArrowLine(160,60)(240,60)
\Vertex(160,60) {1.5}
\ArrowArc(120,60)(40,180,0)
\DashCArc(120,60)(40,0,180) 5
\Text(40,70)[c]{$p$}
\Text(120,110)[c]{$k$}
\Text(120,10)[c]{$p+k$}
\Text(200,70)[c]{$p$}
\end{picture}
\end{center}
\end{minipage}
\caption{The one-loop contribution to the fermion self-energy.}
\end{figure}
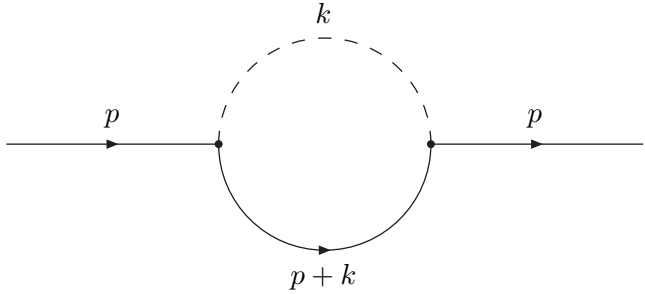
In the framework of the hard thermal loop approximation one then neglects the
external momentum squared, $p^2$, as well as the masses of the particles in the
loop compared to the loop-momentum squared.
After these simplifications the integrations may be performed and one obtains
for the two-point function of a massless fermion ($P=\sqrt{\vec{p}^2}$) 
\bea
(\Delta^f)^{-1}(p) &=& (1+a(p_0,P)) \slash p + b(p_0,P) \slash u \nonumber \\
a(p_0,P) &=& \frac{m_T^2}{P^2} \left[ 1 - \frac{p_0}{2 P} \ln
\left(\frac{p_0+P}{p_0-P}\right)\right] \nonumber \\
b(p_0,P) &=& \frac{m_T^2}{P} \left[ -\frac{p_0}{P} + \frac{1}{2} \left(
\frac{p_0^2}{P^2} -1 \right) \ln
\left(\frac{p_0+P}{p_0-P}\right)\right]
\label{4.17}
\eea
where $m_T^2$ is a quantity with the dimension of (mass)$^2$, given by
\bea
m_T^2 = \frac{\bar{h}^2}{16} T^2 \; .
\label{4.18}
\eea
The dispersion relation corresponding to \eqref{4.17} is given by (we use $u =
(1,0,0,0)$ for the rest-frame of the heat-bath as is commonly done)
\bea
0 = D(\omega) = \left[ 1 + a(\omega,P) \right]^2 \left(
\omega^2-P^2 \right) + 2 \left[ 1 + a(\omega,P) \right] b(\omega,P) \omega +
b(\omega,P)^2 \; .
\label{4.19}
\eea
This equation has four solutions $\pm \omega_\pm$ which
can however only be given analytically in limiting cases. 
A general feature is that the solutions $\omega = \pm \omega_+$ have the same
(positive) ratio of helicity over chirality and are thus connected to the usual
solution at vanishing temperature.
The other solutions, $\omega = \pm \omega_-$ have a negative helicity over
chirality ratio and constitute a collective mode termed the "plasmino". 
As $P \rightarrow 0$, both solutions approach each other and
\bea
\omega_\pm(P) {\lower2.5pt\hbox{$\stackrel{\longrightarrow}{{\scriptscriptstyle{m_T \gg P}}}$}} m_T
\pm \frac{P}{3} \; .
\label{4.20}
\eea
In the opposite limit, for $P \gg m_T$, one finds that $\omega_-(P)$
exponentially approaches $P$ with an also exponentially suppressed residue,
whereas $\omega_+(P) \rightarrow P + m_T^2/P$.
The most important observation is however that the complicated
momentum-dependence of the effective fermion propagator as given in
\eqref{4.17} gives rise to a dispersion relation which resembles for small
momenta $P$ that of a massive particle with a mass $\propto T$, while chiral
symmetry is not broken by this behavior.
It is also clear that the result given in \eqref{4.17} may not be reproduced
using a derivative expansion as is routinely done in calculations using
variants of the Wilsonian renormalization group.
Since on the other hand in the framework of the TRG
it is exactly this behavior that guarantees
fermion-decoupling, we need to find a way to incorporate the above results in
an approximation-scheme for the flow-equations.
We now turn to this question.

The canonical way to incorporate nontrivial momentum-dependence of $n$-point
functions in the renormalization-group approach would be to expand the
effective action in powers of the fields rather than in powers of derivatives.
In order to obtain a closed system of flow-equations one then has to truncate
the infinite tower of $n$-point functions by hand, either by setting the
$n$-point functions with $n$ larger than some $n_{\mathrm{max}}$ to zero, or by
using e.g. perturbative expressions for those.
We could in principle follow such an approach, using scalar $n$-point functions
derived from an effective action as given above and not imposing any
restrictions on the momentum-dependence of the fermion two-point function.
Since however the scale-dependence of the fermion self-energy is given by a
flow-equation which involves a loop integration over a fermion propagator (it
is basically given by the $\Lambda$-derivative of the diagram in figure 1
where the two-point functions are replaced by the modified two-point functions
as introduced in section 2), we would end up with a rather complicated
integro-differential equation.

On the other hand we may also use the knowledge about the structure of
the leading thermal corrections to the fermion propagator as discussed above.
This structure is not changed by the presence of an infrared-cutoff scale
$\Lambda$. 
All dependence of the leading correction on $\Lambda$ is in the thermal
mass-term \eqref{4.18}, which becomes a function of $\Lambda$. 
Thus instead of \eqref{4.17}, the thermal fermion two-point function for finite
values of $\Lambda$ reads to leading order in $T$, taking into account a
possible chiral symmetry-breaking mass $m_{\chi SB}$ now,
\bea
(\Delta^f)^{-1}(p;\Lambda) &=& (1+a(p_0,P;\Lambda)) \slash p +
b(p_0,P;\Lambda) \slash u - m_{\chi SB}\nonumber \\
a(p_0,P;\Lambda) &=& \frac{m_{T,\Lambda}^2}{P^2} \left[ 1 - \frac{p_0}{2 P} \ln
\left(\frac{p_0+P}{p_0-P}\right)\right] \nonumber \\
b(p_0,P;\Lambda) &=& \frac{m_{T,\Lambda}^2}{P} \left[ -\frac{p_0}{P} +
\frac{1}{2} \left( \frac{p_0^2}{P^2} -1 \right) \ln
\left(\frac{p_0+P}{p_0-P}\right)\right] \; .
\label{4.21}
\eea
Correspondingly, the dispersion relation is obtained from \eqref{4.19} by
replacing $m_T$ by $m_{T,\Lambda}$ and subtracting $m_{\chi SB}^2$.
The solutions have the same general features as discussed above. 
If we use this approximation for the fermion propagator in the flow-equations
for the effective potential and the Yukawa-coupling, we find the following
result:
\bea
\Lambda \partial_\Lambda U'(\rho) &=& - \frac{\Lambda^3}{4\pi^2} \left[ \left(
3 U''(\rho) + 2 \rho U^{(3)}(\rho) \right)
\frac{N(\bar{\omega}_\sigma)}{\bar{\omega}_\sigma} + 3 U''(\rho)
\frac{N(\bar{\omega}_\pi)}{\bar{\omega}_\pi} \right] - \nonumber \\
&&
- 2 N_c \frac{\Lambda^3}{\pi^2} \bar{h}^2
\left[ \frac{\tilde{N}(\bar{\omega}_+)}{D'(\bar{\omega}_+)} -
\frac{\tilde{N}(\bar{\omega}_-)}{D'(\bar{\omega}_-)} \right]
\label{4.22}
\eea
and
\bea
\Lambda \partial_\Lambda \bar{h} &=& - \frac{\Lambda^3}{4 \pi^2} \bar{h}^3
\left\{ \frac{N(\bar{\omega}_0)}{\bar{\omega}_0 D(\bar{\omega}_0)} +
2 \left[ \frac{\tilde{N}(\omega_{+,0})}{D'(\bar{\omega}_{+,0}) \left(
\bar{\omega}_{+,0}^2-\bar{\omega}_0^2 \right)} -
\frac{\tilde{N}(\omega_{-,0})}{D'(\bar{\omega}_{-,0}) \left(
\bar{\omega}_{-,0}^2-\bar{\omega}_0^2 \right)} \right] \right\} \; .
\label{4.23}
\eea
Here $\bar{\omega}_\sigma$, $\bar{\omega}_\pi$, and $\bar{\omega}_0$ have been
given in \eqref{3.13} and \eqref{4.3}, whereas $\omega_\pm$ and $\omega_{\pm,0}$ are the positive
solutions of 
\bea
0 = D(\omega) = \left[ 1 + a(\omega,\Lambda;\Lambda) \right]^2 \left(
\omega^2-\Lambda^2 \right) + 2 \left[ 1 + a(\omega,\Lambda;\Lambda) \right]
b(\omega,\Lambda; \Lambda) \omega +
b(\omega,\Lambda; \Lambda)^2 - \frac{\bar{h}^2 \rho}{2}
\label{4.24}
\eea
at $\rho \neq 0$ and $\rho = 0$ respectively. 
Also primes on $D(\omega)$ denote the derivative of $D(\omega)$ from
\eqref{4.24} with respect to $\omega$.
We now obtain an additional flow-equation which governs the dependence of the
thermal mass $m_{T,\Lambda}$ (appearing in $a$ and $b$ above, see \eqref{4.21})
on $\Lambda$. 
This flow-equation reads
\bea
\Lambda \partial_\Lambda m_{T,\Lambda}^2 = - \bar{h}^2 \frac{\Lambda^3}{4
\pi^2} \left\{ \frac{N(\bar{\omega}_0)}{\bar{\omega}_0} + 
2 \left[ \frac{\tilde{N}(\bar{\omega}_{+,0})}{D'(\bar{\omega}_{+,0})}-
\frac{\tilde{N}(\bar{\omega}_{-,0})}{D'(\bar{\omega}_{-,0})} \right]
\right\} \; .
\label{4.25}
\eea
In this way we again have a closed system of differential equations which
furthermore -- in contrast to \eqref{4.1} and \eqref{4.2}, being obtained in
lowest order derivative expansion -- contains the most important corrections
induced by thermal fluctuations.
Even though the boundary value for the thermal mass term $m_{T,\Lambda}^2$ for
$\Lambda \rightarrow \infty$ is zero, by \eqref{4.25} for any finite
$\Lambda \geq 0$
and $\bar{h}^2$ one gets a nonvanishing effective thermal mass for the fermions
even for $\rho=0$.
In this way, remembering the discussion above, fermion decoupling is guaranteed
also at the second order phase-transition, since again the "energy-denominator"
measured in units of $\Lambda$ diverges as $\Lambda \rightarrow 0$.

We could thus stop the discussion at this point and use equations \eqref{4.22}
- \eqref{4.25} to discuss the critical behavior of our model.
However, we shall now show that following a proposal made in 
\cite{FlechsigRebhan} (in a different context),
one may furthermore simplify the system of flow-equations without introducing
large errors in interesting quantities.
This further simplification will turn out to be very useful once we
aim to go beyond the lowest order in the derivative expansion in the
scalar sector. 

The key observation is that one may greatly simplify the effective fermion
propagator without changing the most important features by writing
\bea
\Delta^f(p) &=& \frac{(1+a(p_0,P)) \slash p + b(p_0,P) \slash u + m_{\chi
SB}}{(1+a(p_0,P))^2 p^2 + 2 (1+a(p_0,P)) b(p_0,P) p_0 + b(p_0,P)^2 - m_{\chi
SB}^2} \nonumber \\
&=& \frac{\slash p + m_{\chi SB} + {\mathcal{O}}(m_T^2/P)}{ p^2 - 2 m_T^2 -
m_{\chi SB}^2 + {\mathcal{O}}(m_T^4/P^2)}
\label{4.26}
\eea
and neglecting terms suppressed by powers of $P$. 
Clearly, for $m_{\chi SB}=0$, the resulting expression preserves chiral
invariance on the one hand, but displays the behavior of a massive propagator
as far as the dispersion relation is concerned on the other hand.
If we use this modification, the flow-equations for the potential, the
Yukawa-coupling, and the thermal mass read
\bea
\Lambda \partial_\Lambda U'(\rho) &=& - \frac{\Lambda^3}{4\pi^2} \left[ \left(
3 U''(\rho) + 2 \rho U^{(3)}(\rho) \right)
\frac{N(\bar{\omega}_\sigma)}{\bar{\omega}_\sigma} + 3 U''(\rho)
\frac{N(\bar{\omega}_\pi)}{\bar{\omega}_\pi} \right] -
N_c \frac{\Lambda^3}{\pi^2} \bar{h}^2
\frac{\tilde{N}(\bar{\omega}_\psi)}{\bar{\omega}_\psi} \nonumber \\
\Lambda \partial_\Lambda \bar{h} &=& \frac{\Lambda^3}{4 \pi^2} \bar{h}^3
\frac{1}{U'(0)-2 m_{T,\Lambda}^2} \left[ \frac{N(\bar{\omega}_0)}{\bar{\omega}_0}
+ \frac{\tilde{N}(\bar{\omega}_{\psi,0})}{\bar{\omega}_{\psi,0}} \right]
\label{4.27}
\eea
and
\bea
\Lambda \partial_\Lambda m_{T,\Lambda}^2 = - \bar{h}^2 \frac{\Lambda^3}{4
\pi^2} \left[ \frac{N(\bar{\omega}_0)}{\bar{\omega}_0} + 
\frac{\tilde{N}(\bar{\omega}_{\psi,0})}{\bar{\omega}_{\psi,0}}\right]
\label{4.28}
\eea
with $\bar{\omega}_\psi^2 = \Lambda^2 + 2 m_{T,\Lambda}^2 + m_{\chi SB}^2$,
$\bar{\omega}_{\psi,0}^2 = \Lambda^2 + 2 m_{T,\Lambda}^2$.
Since the fermions decouple from the universal critical behavior both in the
simplified version of the flow-equations (\eqref{4.27} and \eqref{4.28}) as
well as in the full HTL-improved versions \eqref{4.22}-\eqref{4.25}, all
universal results will be independent of the simplifications performed in order
to arrive at \eqref{4.27}.

\begin{figure}[t]
\centering\epsfig{file=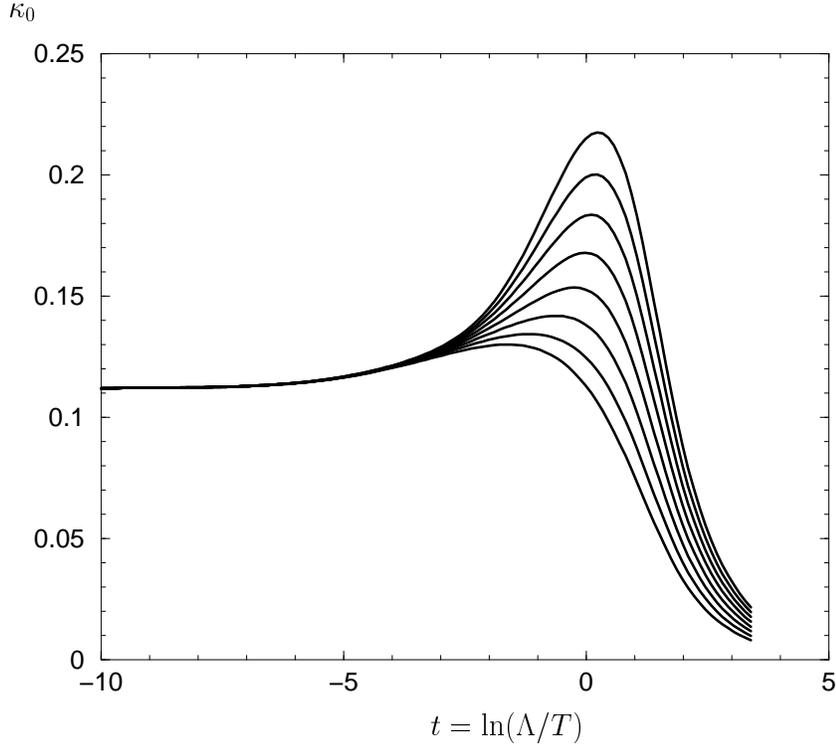,width=12cm}
\caption{The dimensionless minimum $\kappa_0$ as a function of $t$ at the
critical temperature for various values of $x =
\frac{\bar{h}^2_{(T=0)}}{g_{(T=0)}}$ ($x=0.01,0.2,0.4,0.6,0.8,1,1.2,1.4$ from
lowest to uppermost curve).}
\end{figure}

\begin{figure}[t]
\centering\epsfig{file=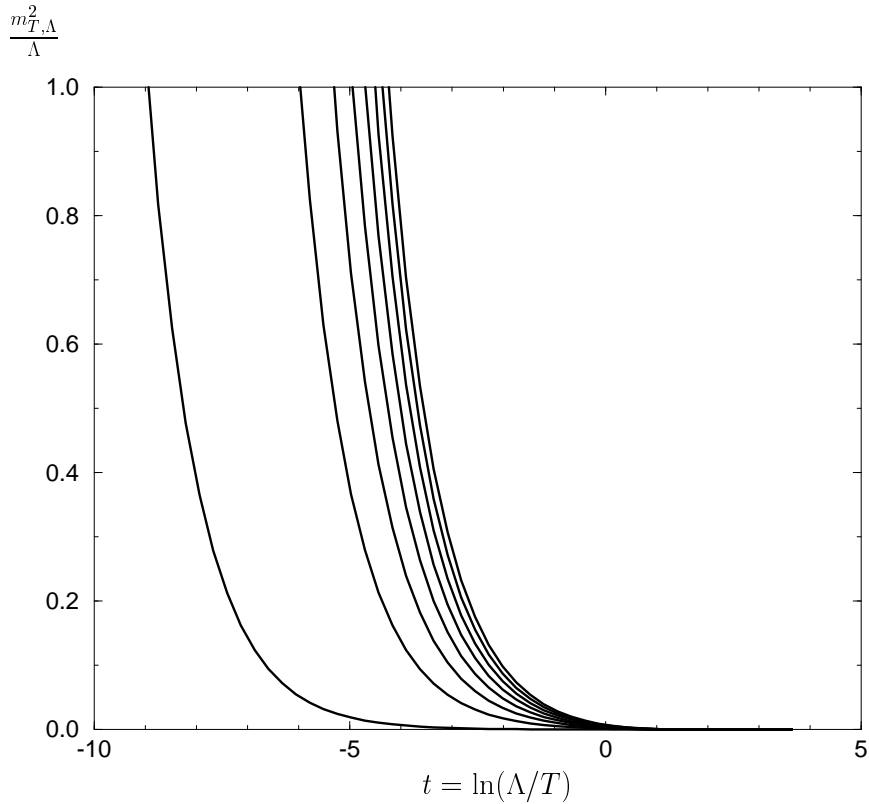,width=12cm}
\caption{The value of the thermal mass $m_{T,\Lambda}^2$ divided by $\Lambda$
as a function of $t$ at the
critical temperature for various values of $x$ ($x=0.01,0.2,0.4,0.6,0.8,1,1.2,1.4$ from
leftmost to rightmost curve).}
\end{figure}

Before we turn to a discussion of the universal critical behavior in the next
section, let us consider the modifications brought about by the
emergence of a thermal mass for the fermions and the quality of the
approximation \eqref{4.26}.
In order to check the universality of the critical behavior we may consider
the model for different values of the Yukawa coupling $\bar{h}_{(T=0)}$ at the
respective critical temperature.
If we approach the critical temperature from below, the effective potential
always has a nontrivial minimum at a finite value of $\kappa = \kappa_0$.
If we consider the rescaled field and couplings according to \eqref{4.6}, we
expect the value of $\kappa_0$ to be independent of $\lambda$ for small values
of the flow-parameter.
In figure 2 we display the dependence of the rescaled minimum of the potential
on $t=\ln \lambda$ at the critical temperature for various initial values of
the Yukawa coupling (unless stated otherwise, we always choose
$g_{(T=0)} = 0.1$ 
and vary $x = \bar{h}^2_{(T=0)}/g_{(T=0)}$).
A scaling solution is reached for $t \klgl -7$ and one nicely observes the
independence of the final value $\kappa_{0,\star}$ of $x$. 
It is remarkable that the fermion-contribution is sizable down to $t < 0$,
i.e.~$\Lambda < T$.

In order to discuss the importance of contributions neglected according to
\eqref{4.26} we display in figure 3 the value of the thermal fermion-mass
$m_{T,\Lambda}^2$ divided by $\Lambda$ (all dimensionful quantities are given
in units of $\sqrt{\rho_{(T=0)}}$).
This is the relevant quantity in applications of \eqref{4.26} in the framework
of the TRG with a sharp cutoff, since the three-momenta inside the loop
contributing to the flow of any quantity are $P=\Lambda$ due to the
$\delta$-function in the kernel (\eqref{kernelscalar} and correspondingly for
the fermionic degrees of freedom).
We plot in figure 3 $m_{T,\Lambda}^2/\Lambda$ again as a function of $t$ for
the same values of $g_{(T=0)}$ and $x$ as used in figure 2. 
We note that the thermal mass remains very small for large $t$ -- a one-loop
calculation for finite $\Lambda$ yields 
\bea
m_{T,\Lambda}^2 &=& \int_\Lambda^\infty dx \bar{h}^2 \frac{x}{4\pi^2} \left( \frac{1}{e^{x/T}-1} + \frac{1}{e^{x/T}+1} \right) \nonumber \\
&=& \frac{\bar{h}^2}{12} T^2 + \frac{\bar{h}^2}{4\pi^2} T \left[ T {\mathrm{dilog}}( e^{\Lambda/T}) + T {\mathrm{dilog}} ( e^{\Lambda/T}+1 ) + \Lambda \ln ( e^{\Lambda/T}+1 ) \right] \nonumber \\
&\approx& \frac{\bar{h}^2}{2\pi^2} T e^{-\Lambda/T} (\Lambda+T) \quad ({\mbox{$\Lambda/T$ large}}) 
\label{4.29}
\eea
and $m_{T,\Lambda}^2/\Lambda$ is smaller than $0.2$ for the values of $x$
considered here for all $t \grgl -2.5$.
On the other hand, as is obvious from figure 2, the main contribution from the
fermionic degrees of freedom is at larger values of $t$.
Thus we believe that the simplification of the fermionic dispersion relation
made here introduces only very small errors for the nonuniversal results -- as
was pointed out above, universal quantities are completely
unaffected\footnote{Also note that neglected contributions in \eqref{4.26} are
formally of higher order in $\bar{h}^2$ through \eqref{4.29}.}.

\begin{figure}[t]
\centering\epsfig{file=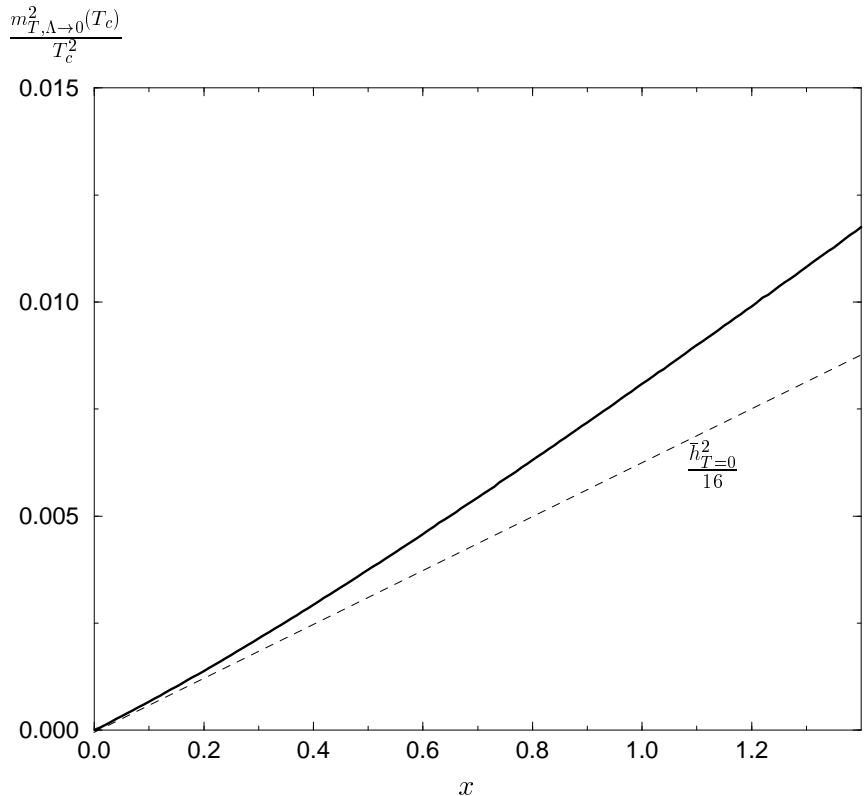,width=12cm}
\caption{The thermal mass $m_{T,\Lambda\rightarrow 0}^2$ in units of $T^2$ at
the critical temperature as a function of $x$ (solid line). 
For comparison the leading order result from \eqref{4.18} is also given (dashed
line).}
\end{figure}

Before turning to the phase-transition itself, consider finally the value of
$m_{T,\Lambda}^2$ as $\Lambda \rightarrow 0$.
This is displayed in figure 4, where we have plotted $m_{T,\Lambda \rightarrow
0}^2/T^2$ at the critical temperature as a function of $x$ again.
For comparison the one-loop result \eqref{4.18} is displayed as dashed line.
The renormalization of the Yukawa-coupling through thermal fluctuations yields
a small but non-negligible deviation of this quantity from its lowest-order
value.
We point out that here the HTL-contribution (in the fermionic sector
approximated by \eqref{4.26}) is consistently resummed.

\section{The universal critical behavior at the chiral phase-transition}

In this section we finally discuss the results for the chiral phase-transition
as obtained from the quark-meson model after including the leading thermal
effects in the fermionic sector as discussed above.
We have pointed out that the derivative expansion fails for the fermion
propagator and that fermion decoupling at a second order phase-transition is
only guaranteed by the nonlocal fermion mass-term produced by the thermal
fluctuations. 
After taking this effect into account, we expect the phase-transition to be in
the universality class of the $O(4)$-model. 
Critical exponents and amplitude-ratios should be independent of the number of
colors or the Yukawa-coupling.
Before we turn to the scalar sector and the universal quantities, let us
briefly discuss two nonuniversal aspects: the critical temperature and the
behavior of the Yukawa-coupling.

\begin{figure}[t]
\centering\epsfig{file=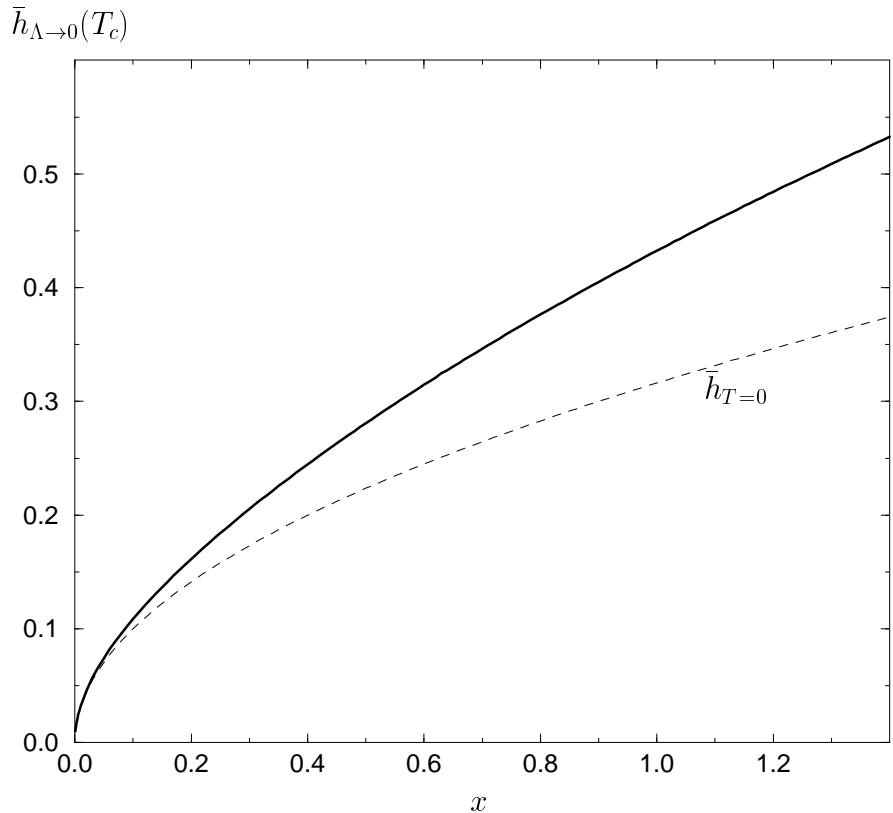,width=12cm}
\caption{The Yukawa-coupling $\bar{h}_{\Lambda\rightarrow 0}(T_c)$ as a function of $x$ (solid line). 
For comparison $\bar{h}_{(T=0)}$ is also given (dashed line).}
\end{figure}

In this section we introduce an additional effect, namely we go to first order
in the derivative expansion in the scalar sector by taking into account a
scalar wave-function renormalization.
This quantity was already included in $\bar{\Gamma}_\Lambda$,
\eqref{gammaqueron}.
Whereas $Z_\Lambda$ is in principle a function of the order-parameter field, we
evaluate it at the minimum of the effective potential.
Its $t$-dependence is given by
\begin{equation}
\frac{\partial Z_\Lambda}{\partial t} = - \eta Z_\Lambda
\label{5.1}
\end{equation}
with the anomalous dimension $\eta$.
Having introduced $Z_\Lambda$ we consider instead of the dimensionless
quantities from \eqref{4.7} renormalized dimensionless couplings
according to 
\begin{equation}
\kappa_r = Z_\Lambda \kappa \quad ; \quad h^2 = Z_\Lambda^{-1}
\bar{h}^2 \quad ; \quad u_r^{(n)} = Z_\Lambda^{-n} u^{(n)} \; .
\label{5.3}
\end{equation}

We evaluate the anomalous dimension by considering the flow-equation for the
$\pi$-two-point function and expanding it around $\vec{q}^2 = 0$.
Since the heat-bath introduces an additional four-vector $u_\mu$, there are in
principle two different wave-function renormalizations for (choosing as usually
$u_\mu = (1,0,0,0)$) $\vec{q}$ and $q_0$.
We do not make this distinction and consider only the effect of $Z_\Lambda$
connected to $\vec{q}^2$, since this is the quantity which through \eqref{5.1}
yields the critical exponent $\eta$ (see below).
For the anomalous dimension we find the following expression:
\bea
\eta &=& \left.\lambda \frac{\kappa_r \big ( u_r^{(2)}
\big)^2}{12\pi^2}\bigg(F(\omega_{\sigma},\omega_{\pi})+F(\omega_{\pi},\omega_{\sigma})\bigg)\right|_{\kappa_r=\kappa_{r,0}}
\nonumber\\
&& +N_c \frac{h^2}{12 \pi^2 \omega_\psi^7} 
\bigg \{ \big[6\omega_\psi^4 - 3\omega_\psi^2 -
9\mu_T^2\omega_\psi^2 + 10 \mu_T^2\big]
\big({-\tilde{n}}(\lambda\omega_\psi) + \lambda \omega_\psi
\tilde{n}(\lambda\omega_\psi) (\tilde{n}(\lambda\omega_\psi)
- 1)\big ) + \nonumber \\
&& \hspace{2.5cm}\big [ \omega_\psi^2 + 3 \omega_\psi^2\mu_T^2 -4
\mu_T^2\big ]
\lambda^2 \omega_\psi^2 \tilde{n}(\lambda\omega_\psi) \big (
\tilde{n}(\lambda\omega_\psi) - 1 \big ) \big (
2\tilde{n}(\lambda\omega_\psi) - 1 \big ) + \nonumber \\
&&\hspace{2.5cm}\frac{2}{3} \mu_T^2 \lambda^3 \omega_\psi^3
\tilde{n}(\lambda\omega_\psi)\big
(\tilde{n}(\lambda\omega_\psi) - 1\big) \big( 6
\tilde{n}(\lambda\omega_\psi)^2 - 6\tilde{n}(\lambda\omega_\psi)  +
1\big)  \bigg \}
\label{5.2}
\eea
with
\bea
F(\omega_{\sigma},\omega_{\pi})&=&
\frac{1}{\omega_{\sigma}\omega_{\pi}^{5}(\omega_{\sigma}^2-\omega_{\pi}^2)^{3}}\bigg
\lbrace
n(\lambda\omega_{\sigma})(6\omega_{\pi}^7-8\omega_{\pi}^5-6\omega_{\sigma}^2\omega_{\pi}^5) 
\nonumber\\ 
& &
+n(\lambda\omega_{\pi})(3\omega_{\sigma}^5-3\omega_{\sigma}^5\omega_{\pi}^2-10\omega_{\sigma}^3\omega_{\pi}^2+12\omega_{\sigma}^3\omega_{\pi}^4+15\omega_{\sigma}\omega_{\pi}^4-9\omega_{\sigma}\omega_{\pi}^6)\nonumber\\
& & -\lambda
n(\lambda\omega_{\pi})(n(\lambda\omega_{\pi})+1)(3
\omega_{\sigma}^5\omega_{\pi}^3-3\omega_{\sigma}^5\omega_{\pi}+10\omega_{\sigma}^3\omega_{\pi}^3-6\omega_{\sigma}^3\omega_{\pi}^5+3\omega_{\sigma}\omega_{\pi}^7-7\omega_{\sigma}\omega_{\pi}^5)\nonumber\\
& &+\lambda^2
n(\lambda\omega_{\pi})(n(\lambda\omega_{\pi})+1)(2n(\lambda\omega_{\pi})+1)(\omega_{\sigma}^5\omega_{\pi}^2-2\omega_{\sigma}^3\omega_{\pi}^4+\omega_{\sigma}\omega_{\pi}^6)\,\,\bigg
\rbrace
\label{5.2.1}
\eea
where $\mu^2_T = \frac{m_{T,\Lambda}^2}{\Lambda^2}$, $\mu_\chi^2 = \frac{h^2 \kappa_{r,0}}{2 \lambda}$, $\omega_\psi^2 = 1 + 2 \mu_T^2 + \mu_\chi^2$ and $\omega_\sigma$ as well as $\omega_\pi$ are defined in \eqref{4.10} (where the renormalized quantities according to \eqref{5.3} should be used).
In the limit where $\Lambda$ is small compared to all other scales in the theory (this is the limit relevant for the fixed-point behavior) this rather complicated expression reduces to
\bea
\eta = \left.\frac{\kappa_{r}\big ( u_r^{(2)}\big)^2 }{6
\pi^2}\frac{4\omega_{\sigma}^4-3\omega_{\sigma}^4\omega_{\pi}^2-3\omega_{\sigma}^2\omega_{\pi}^4+4\omega_{\pi}^4}{\omega_{\sigma}^6\omega_{\pi}^6} 
\right|_{\kappa_r=\kappa_{r,0}} + {\mathcal{O}}(\lambda)
\label{5.2.a}
\eea
and we again note the decoupling of the fermionic degrees of freedom in this limit.

The flow-equations for the renormalized couplings introduced in \eqref{5.3} receive additional contributions
proportional the anomalous dimension.
We thus have for example
\bea
\partial_t h &=& \frac{\eta}{2} h + \frac{h^3}{4\pi^2 u_r^{(1)}(0)} \left[ \frac{n(\lambda \omega_{r,0})}{\omega_{r,0}} + \tilde{n}(1) \right] \nonumber \\
\partial_t u_r^{(1)}(\kappa_r) &=& (-2+\eta) u_r^{(1)} + (1+\eta) \kappa_r u_r^{(2)}- \nonumber \\
&& \quad - \frac{\lambda}{4\pi^2} \left[ \left( 3 u_r^{(2)} + 2 \kappa_r u_r^{(3)} \right) \frac{n(\lambda \omega_\sigma)}{\omega_\sigma} + 3 u_r^{(2)} \frac{n(\lambda \omega_\pi)}{\omega_\pi} \right] - \frac{N_c}{\pi^2} h^2 \frac{\tilde{n}(\lambda \omega_\psi)}{\omega_\psi}
\label{5.4}
\eea
and for the renormalized dimensionless minimum 
\bea
\partial_t \kappa_{r,0} = -(1+\eta) \kappa_{r,0} + \left. \frac{\lambda}{4\pi^2} \left[ \left( 3 + 2 \kappa_r \frac{u_r^{(3)}}{u_r^{(2)}} \right) \frac{n(\lambda \omega_\sigma)}{\omega_\sigma} + 3 \frac{n(\lambda \omega_\pi)}{\omega_\pi} \right] \right|_{\kappa_r = \kappa_{r,0}} + \left. \frac{N_c}{\pi^2} \frac{h^2}{ u_r^{(2)} } \frac{\tilde{n}(\lambda \omega_\psi)}{\omega_\psi} \right|_{\kappa_r = \kappa_{r,0}}
\label{5.5}
\eea
Since the value of the anomalous dimension at the critical temperature will
turn out to be very small, the modifications of the results displayed in
figures 3 and 4 are tiny.

The value of the unrenormalized Yukawa-coupling $\bar{h}^2$ at $T_c$ as
$\Lambda \rightarrow 0$ is displayed as a function of $x$ in figure 5.
Also plotted is the $(T\!=\!0)$-Yukawa-coupling and we note that it is
only slightly 
modified through thermal fluctuations (for small values of the initial
couplings).
Here the introduction of a renormalized Yukawa-coupling however has important
consequences: The value of $\eta$ is positive at the critical temperature. 
This leads to a vanishing renormalized Yukawa-coupling at $T_c$, since in the
infrared the flow-equation for $h^2$ reduces to
\begin{equation}
\partial _t h^2 = \eta h^2 \qquad ({\mbox{$T=T_c, \Lambda \rightarrow
0$}}) \; .
\label{5.6}
\end{equation}
Since $\eta$ is very small, $h^2$ only vanishes very slowly ($h^2 \propto
\lambda^\eta$).

\begin{figure}[t]
\centering\epsfig{file=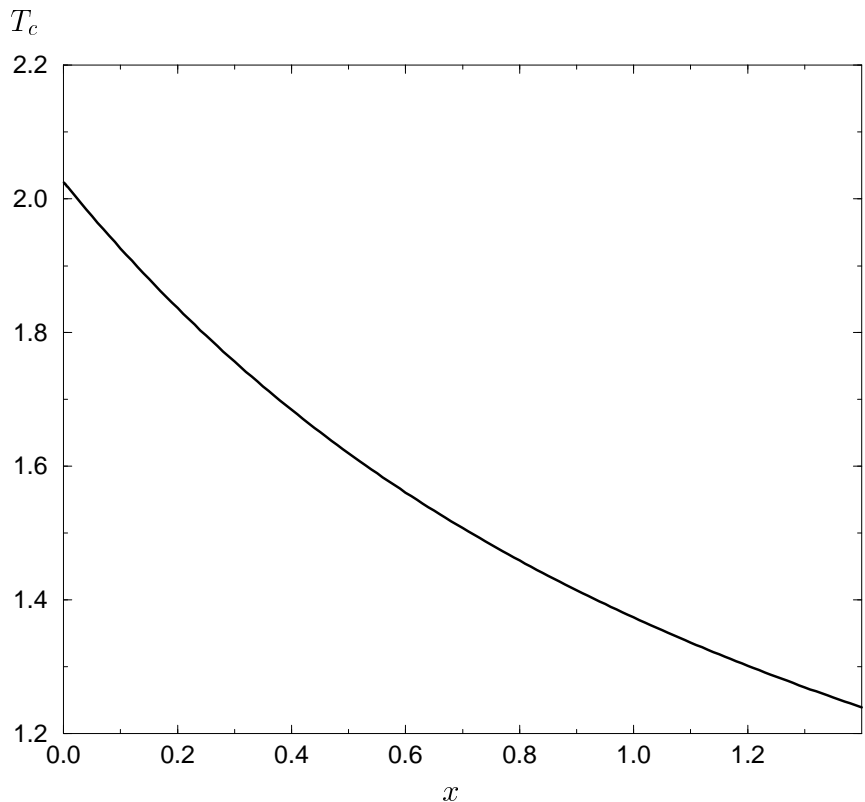,width=12cm}
\caption{The critical temperature in units of $\sqrt{\rho_0(T=0)}$ as a
function of $x = \frac{\bar{h}^2_{(T=0)}}{g_{(T=0)}}$ (we use
$g_{(T=0)} = 0.1$ here).} 
\end{figure}

Let us now turn to the critical temperature.
In leading order perturbation theory, one obtains $T_c$ as a function of $x =
\bar{h}^2_{(T=0)}/g_{(T=0)}$ as
\bea
T_c^{\mathrm{pert}} = \sqrt{\frac{4 \rho_{0}(T=0)}{1+x}} \; .
\label{5.7}
\eea
This dependence on $x$ is basically reproduced by our resummed calculation.
We display the critical temperature in units of $\sqrt{\rho_{0}(T=0)}$ as a
function of $x$ in figure 6.
Note that in the limit $x \rightarrow 0$ the result does not approach
\eqref{5.7}.
This is of course due to the fact that this limit corresponds to a pure scalar
theory where there are still modifications of $T_c^{\mathrm{pert}}$ 
\cite{BJ1}.
\begin{figure}[t]
\centering\epsfig{file=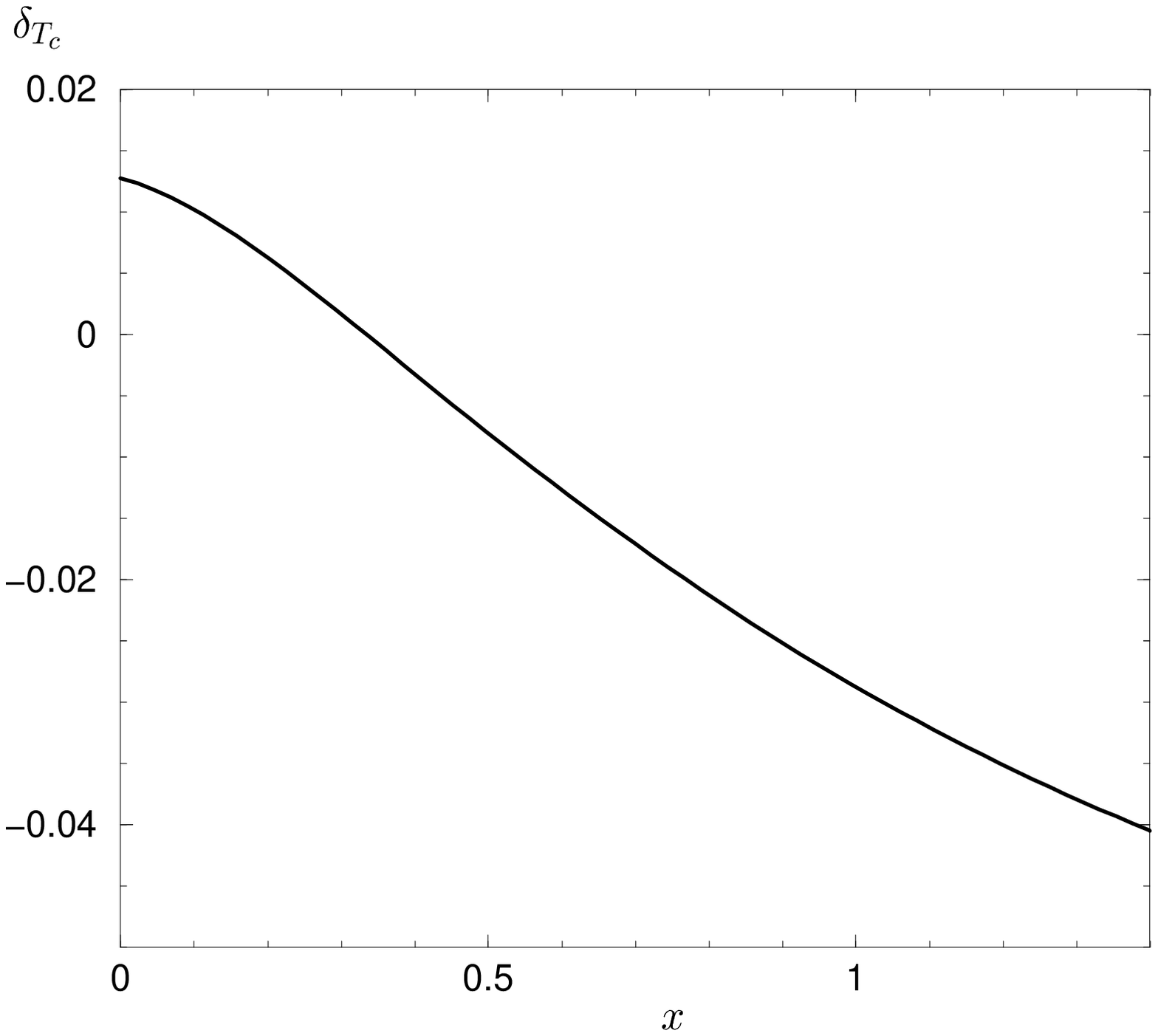,width=12cm}
\caption{The relative deviation of the critical temperature from the
perturbative result, $\delta_{T_c} = 2
\frac{T_c-T_c^{\mathrm{pert}}}{T_c+T_c^{\mathrm{pert}}}$ as a function of $x$ ($g_{(T=0)} = 0.1$).}
\end{figure}
Indeed, the relative difference of our result with lowest order perturbation
theory as displayed in figure 7 is small but nonvanishing for $x
\rightarrow 0$.

To end the discussion, let us report the results for the universal behavior.
As the universal quantities of the $O(N)$-models are well studied for various
values of $N$ (for a review see 
\cite{ZinnJustin})
and the cases $N=1$ and $N=4$ have been studied with the methods used here in
\cite{DAP1,B1,BJ1}
we will only briefly discuss the main differences of our results with the ones
that may be found in the literature.

The universal critical behavior is encoded in the scaling form of the equation
of state, which relates the order parameter $\rho$, the temperature $T$ and the
external field $H = \sqrt{2 \rho} U'(\rho)$ and may for example be written as
\cite{ZinnJustin}
\bea
\frac{\partial U}{\partial \rho} = \left( \sqrt{ 2 \rho} \right)^{\delta - 1}
f(x) \qquad ; \qquad x = \frac{T-T_c}{T} \left( \sqrt{ 2 \rho}
\right)^{-1/\beta}
\label{5.8}
\eea
with the Widom scaling function $f(x)$ and critical exponents $\delta$ and
$\beta$.
The Widom scaling function for the $O(4)$-model was calculated in the framework
of the thermal renormalization group in 
\cite{BJ1}, 
where however the anomalous dimension was neglected.
It has also been computed from the exact renormalization group in the
Matsubara-formalism 
\cite{JNC2},
from three-dimensional lattice simulations
\cite{Toussaint},
and from mean-field theory as well as in second order $\epsilon$-expansion
\cite{BrezinWallaceWilson}.
The TRG-results have been compared to those of the exact renormalization group
in
\cite{BJ1}
and good agreement was found apart from the behavior for large $x$ and
$T>T_c$.
The disagreement in that region was due to differences in the value of the
critical exponent $\gamma$, which governs the behavior of $f(x)$ as $x
\rightarrow \infty$. In 
\cite{BJ1} 
it was argued that these differences should mostly be due to the neglected
effects of the anomalous dimension.
We are now in a position to check this claim through a calculation of critical
exponents taking into account $\eta$ according to \eqref{5.2}.

We have calculated the exponents $\beta$ and $\delta$ as introduced in
\eqref{5.8}, $\eta$ as given in \eqref{5.2}, as well as $\gamma$ and $\nu$.
The exponents $\gamma$ and $\nu$ encode the behavior of the unrenormalized and
the renormalized mass of the order parameter field as $T \rightarrow T_c$
according to
\bea
\frac{m_r(T)}{T_c} \propto \left( \frac{T-T_c}{T_c} \right)^\nu \nonumber \\
\frac{m(T)}{T_c} \propto \left( \frac{T-T_c}{T_c} \right)^{\gamma/2}
\label{5.9}
\eea
whereas $\eta$ describes the behavior of the two-point function of the
order-parameter field at the critical temperature.
Our results for these exponents are given in table 1, where we also give the
results obtained from other methods for comparison\footnote{For results from 
the ERG see also \cite{NC1} where results that are essentially identical 
to those given in \cite{JNC2} where obtained from a polynomial approximation 
to the effective potential.}.

A useful consistency check is provided by the scaling relations which connect
the critical exponents and leave only two of them independent.
Thus for example one has
\cite{ZinnJustin}
\bea
\delta=\frac{d+2-\eta}{d-2+\eta} \nonumber \\
\gamma = \nu (2 - \eta) \nonumber \\
\gamma = \beta (\delta - 1) \nonumber \\ 
\beta = \frac{\nu}{2} (d - 2 + \eta)  
\label{5.10}
\eea
These scaling relations are fulfilled by our results to better than $0.5 \%$.

\setlength{\arrayrulewidth}{0.25mm}
\setlength{\doublerulesep}{0.25mm}
\begin{table}[t]
\begin{center}
\begin{tabular}{||c||c|c|c|c|c||}
\hline \hline
  & $\beta$ & $\gamma$ & $\nu$ & $\delta$ & $\eta$ \\
\hline \hline
This work & $0.429$ & $1.68$ & $0.85$ & $4.90$ & $0.017$ \\
\hline
TRG + LO DE \cite{BJ1} & $0.433$ & $1.73$ & $0.86$ & $5.0$ & - \\
\hline 
\hline
ERG \cite{JNC2} & $0.407$ & $1.548$ & $0.787$ & $4.80$ & $0.0344$ \\
\hline
3d PT \cite{Bakeretal} & $0.38$ & $1.44$ & $0.73$ & $4.82$ & $0.03$ \\
\hline
3d MC \cite{KanayaKaya} & $0.384$ & $1.48$ & $0.748$ & $4.85$ & $0.025$ \\
\hline \hline
\end{tabular}
\end{center}
\caption{Critical Exponents for the $O(4)$ model.}
\label{crit_exp}
\end{table}

The results from \cite{Bakeretal} are obtained from a fixed-dimension
calculation in the three-dimensional $O(4)$-model to seven loops whereas the
results given in the last line are from three-dimensional Monte-Carlo
simulations of the $O(4)$-model \cite{KanayaKaya}.
We note that the results obtained in the present work, derived using the
next-to-leading contributions in the derivative expansion in the scalar sector,
are somewhat closer to the values found in lattice and three-dimensional perturbation theory-calculations than the results given in \cite{BJ1}.
However, the improvement for the exponents $\beta$, $\gamma$ and $\nu$ is perhaps surprisingly small for the present model.

At this point it is interesting to compare the situation with the results from a study of the pure scalar $O(1)$-model.
In 
\cite{SenBenMichael}
a sharp-cutoff Wilsonian RG was used to obtain critical exponents of the three-dimensional $O(1)$-model.
We have explicitely verified that our flow-equation for the effective potential and the anomalous dimension, adapted to the $O(1)$ scalar theory\footnote{This is trivial for the flow-equation for the potential. The anomalous dimension has to be replace by the one defined from the Higgs since there are of course no Pions in the $O(1)$-case. We have calculated the corresponding expression also in the present model.} reproduce the expression given there in the scaling limit.
The truncations made in \cite{SenBenMichael} are then identical to the ones made here and we would reproduce their values for the exponents.
In that case, the exponents from the Wilsonian RG are in very good agreement with the results from other methods.
This shows that in the $O(1)$-model, the next-to-leading order derivative expansion as used here gives accurate results as one would expect given the smallness of the anomalous dimension\footnote{As a caveat, one should however notice that the scaling relations \eqref{5.10} are violated by up to $\sim 6 \%$ for the values given in 
\cite{SenBenMichael}.}.
The fact that the agreement in the present case is not as good might be due to the existence of a contribution in the present order of the derivative expansion which exists for $N>1$ and has been neglected here, namely a term of the form
$\tilde{Z}_\Lambda \varphi_a \varphi_a \partial^2 \varphi_b \varphi_b$
(see also \cite{BJ1}).
This question remains to be settled in future work.

\section{Summary and Conclusions}

In the present work we have extended the formulation of the thermal
renormalization group 
\cite{DAP1}
to theories involving fermionic degrees of freedom.
Whereas the adaption of the full flow-equation (section 2) is very much
straightforward and does not involve any conceptually new points, it
turns out that in performing approximations to the full functional
differential equation one has to be considerably more careful for
fermionic than for bosonic fields. After describing in section 3 the
specific model for the $N_f=2$ chiral phase-transition that we
studied, we have discussed at length questions concerning the
decoupling of the fermionic fields at the phase-transition in section 4.

The model under consideration consists of an $O(4)$-symmetric scalar
field coupled via Yukawa-couplings to $2 N_c$ chiral fermions.
At low temperatures the $O(4)$-symmetry is spontaneously broken down to an
$O(3)$, where correspondingly 3 massless scalar fields ('pions') appear.
The massive scalar field is the would-be 'sigma', the fermionic fields
acquire masses and the chiral symmetry thus is spontaneously broken.
This model is expected to be in the universality class of the
three-dimensional 4-vector model with a second order phase-transition
with well known universal behavior.
The fermionic fields should not play a r{\^o}le for the infrared-behavior
at the phase-transition and the effective potential should display scaling
behavior for small field-values. 

We have then pointed out that the usual arguments about fermion
decoupling at high temperatures do not apply in the TRG-approach in
the presence of second- or weakly first order phase-transitions.
The fact that the scalar distribution function dominates the fermionic
one for small energies (or large temperatures) does not directly imply
a dominance of the scalar contributions to the renormalized couplings
(or the effective potential).
The reason for this is the fact that it is not the distribution
functions alone that enter the calculations, but the product of
couplings times distribution functions. 
In the presence of a scaling solution for the effective potential, the
couplings are also scale dependent in such a way that the product of
scalar self couplings times scalar distribution functions has the same
infrared behavior as the fermionic contribution, since in the latter the
couplings do not exhibit scaling behavior.

This is in contrast to the high-temperature limit, where neither the
scalar nor the Yukawa-couplings scale. 
In this case it is indeed the relative behavior of the distribution
functions that matters and fermion decoupling in the limit
$T\rightarrow \infty$ works by the usual arguments.

One of the main points of the present work was then made in section 4
where we have discussed the question how universality in the presence
of chiral fermions is nevertheless obtained and how this effect can be
successfully implemented within the thermal renormalization
group-approach. The key observation is that even in chirally symmetric
theories thermal fluctuations modify the fermionic dispersion
relations such that the corrected dispersion relation resembles that
of a massive particle for small momenta
\cite{fermionmass}.
This is the leading effect of finite temperature in the fermionic
sector of the theory and is in this sense on equal footing with the
scalar mass-correction $\propto T^2$ which drives the phase-transition.
Due to the chiral nature of the fermions, the thermal mass in the
fermionic sector is however necessarily an effective mass resulting
from a complicated momentum dependence of the two-point function.
We have pointed out that this effect is not included if one applies a
derivative expansion to the fermion two-point function as is generally
done in the application of Wilsonian renormalization group-equations.
While universal quantities are unaffected, the presence of the
effective thermal fermion mass should modify nonuniversal quantities
as obtained within the derivative expansion
\cite{JNC2}.

We also discussed in section 4 how one may approximate the fermionic
dispersion relations such that the resulting expressions are simple
enough to keep the system of flow-equations manageable on the one hand
and not loose the important effects on the other hand.
Here we follow a proposal put forward in 
\cite{FlechsigRebhan}
to rewrite the fermion propagator such that the thermal mass-term is
effectively local, but chiral symmetry remains unbroken by it.
We argue that this approximation is quantitatively reliable for small
$(T\!=\!0)$-couplings.

After having clarified how fermion-decoupling works within the 
TRG we have finally
discussed  the results of a study of the chiral phase-transition in
the framework of the quark-meson model in section 5.
Indeed, the phase-transition is second order with universal critical
behavior as obtained from the pure scalar $O(4)$-model in three
dimensions. We find that for small enough $(T\!=\!0)$-couplings the
critical temperature is rather well described by perturbation theory.
This observation was already made in 
\cite{CN1,BJ1,NC2} 
for pure scalar $O(N)$-models.
We improved upon previous results from the TRG on the critical
exponents by including a scalar wave-function renormalization,
i.e.~going beyond leading order in the derivative expansion in the
scalar sector. The critical exponents are displayed in table 1.
We have also discussed the quality of our approximations as far as the
critical exponents are concerned.
It turns out that the inclusion of the anomalous
dimension has a rather small effect on the exponents $\beta$, $\gamma$ and 
$\nu$.

In the present paper we have been concerned only with quantities which may
also be obtained from Wilsonian renormalization group-calculations in the
Matsubara-formalism.
In principle, the formalism set up here is directly applicable to
quantities like damping rates etc
\cite{Massimo}.
After we have clarified the necessary ingredients in the fermionic
sector, we are now in a position to study out-of-equilibrium
properties of the present model in linear response theory.

\bigskip

{\bf{Acknowledgments:}} B.B. was supported by 
the "Sonder\-for\-schungs\-be\-reich 375-95 f\"ur
Astro-Teilchen\-physik" der Deutschen Forschungsgemeinschaft, J.M. acknowledges
support by the Deutsche Forschungsgemeinschaft and J.R. was supported by a Promotionsstipendium des Freistaates Bayern.

\end{document}